\documentclass[apjl]{emulateapj}
\usepackage{apjfonts,amssymb,amsbsy}
\usepackage{graphicx}
\usepackage{natbib}
\usepackage{color}
\usepackage{ctable}

\newcommand{\eb}[1]{\begin{equation}\label{eq:#1}}
\newcommand{\en}{\end{equation}}
\newcommand{\eab}[1]{\begin{eqnarray}\label{eq:#1}}
\newcommand{\ean}{\end{eqnarray}}

\newcommand{\ddt}[1]{{\partial #1\over\partial t}}

\newcommand{\ddi}[1]{{\partial #1\over\partial r_i}}
\newcommand{\ddj}[1]{{\partial #1\over\partial r_j}}

\newcommand{\grad}[1]{{\nabla #1}}
\renewcommand{\div}{\nabla\cdot}

\newcommand{\curl}{\nabla\times}

\newcommand{\half}{\frac{1}{2}}
\newcommand{\lnr}{\ln \rho}
\newcommand{\uu}{\mathbf{u}}
\newcommand{\uw}{u_{\mbox{w}}}
\newcommand{\uc}{\,\delta u^+}

\newcommand{\ds}{\Delta s\,}
\newcommand{\BB}{\mathbf{B}}
\newcommand{\TT}{\underline{\mathbf{\tau}}}
\newcommand{\rr}{\mathbf{r}}
\newcommand{\ff}{\mathbf{f}}
\newcommand{\ijk}{{i,j,k}}
\newcommand{\Sij}{{S_{ij}}}
\newcommand{\Tij}{{\tau_{ij}}}
\newcommand{\IJK}{{i,j,k}}
\newcommand{\RMS}{\mbox{rms}}
\newcommand{\ft}{\tilde{\mathbf{f}}}
\newcommand{\ut}{\tilde{\mathbf{u}}}
\newcommand{\bin}{{\mbox{bin}}}
\newcommand{\kk}{\mathbf{k}_\IJK}
\newcommand{\ik}{i\kk}

\newcommand{\FT}{{\mbox{FT}}}

\definecolor{ENZO}{rgb}{0,0,0}
\definecolor{STAGGER}{rgb}{0,0,0}
\definecolor{KT-MHD}{rgb}{0,0,0}
\definecolor{LL-MHD}{rgb}{0,0,0}
\definecolor{PLUTO}{rgb}{0,0,0}
\definecolor{PPML}{rgb}{0,0,0}
\definecolor{RAMSES}{rgb}{0.000000, 0.000000, 0.000000}
\definecolor{ZEUS}{rgb}{0,0,0}
\definecolor{FLASH}{rgb}{0,0,0}

\def\ENZO{{\color{ENZO} ENZO}}
\def\STAGGER{{\color{STAGGER} STAGGER}}
\def\KTMHD{{\color{KT-MHD} KT-MHD}}
\def\ZEUS{{\color{ZEUS} ZEUS}}
\def\ENZO{{\color{ENZO} ENZO}}
\def\LLMHD{{\color{LL-MHD} LL-MHD}}
\def\PLUTO{{\color{PLUTO} PLUTO}}
\def\PPML{{\color{PPML} PPML}}
\def\RAMSES{{\color{RAMSES} RAMSES}}
\def\FLASH{{\color{FLASH} FLASH}}

\def\apj{ApJ}
\def\apjs{ApJS}

\def\apjl{ApJL}
\def\araa{ARA\&A}

\def\aap{A\&A}
\def\mnras{MNRAS}
\def\prl{Phys. Rev. Lett.}
\def\pre{Phys. Rev. E}
\def\jcp{J. Comput. Phys.}
\def\lsim{~\raise0.3ex\hbox{$<$}\kern-0.75em{\lower0.65ex\hbox{$\sim$}}~}
\def\gsim{~\raise0.3ex\hbox{$>$}\kern-0.75em{\lower0.65ex\hbox{$\sim$}}~}
 
\slugcomment{ApJ Draft Version \today}
\shorttitle{MHD CODE COMPARISON ON TURBULENCE DECAY}
\shortauthors{KRITSUK ET AL.}

\begin{document}
\title{Comparing Numerical Methods for Isothermal Magnetized Supersonic Turbulence}
\slugcomment{Received 2011 March 28; accepted 2011 May 12; published 2011 June}
\journalinfo{The Astrophysical Journal, 735:1 (17pp), 2011}

\author{Alexei G. Kritsuk,\altaffilmark{1,2} 
\AA ke Nordlund,\altaffilmark{2,3}
David Collins,\altaffilmark{1,2,4}
Paolo Padoan,\altaffilmark{2,5}
Michael L. Norman,\altaffilmark{1,6}\\
Tom Abel,\altaffilmark{2,7}
Robi Banerjee,\altaffilmark{2,8,9} 
Christoph Federrath,\altaffilmark{8,10,11}
Mario Flock,\altaffilmark{10}
Dongwook Lee,\altaffilmark{12}
Pak Shing Li,\altaffilmark{2,13}\\
Wolf-Christian M\"uller,\altaffilmark{14}
Romain Teyssier,\altaffilmark{2,15,16}
Sergey D. Ustyugov,\altaffilmark{17}
Christian Vogel,\altaffilmark{14}
and Hao Xu\altaffilmark{1,4}
}
\altaffiltext{1}{Physics Department and Center for Astrophysics and Space Sciences,
University of California, San Diego; 9500 Gilman Drive, La Jolla, CA 92093-0424, USA; akritsuk@ucsd.edu}

\altaffiltext{2}{Kavli Institute for Theoretical Physics, University of California, Santa Barbara, CA 93106-4030, USA}

\altaffiltext{3}{Centre for Star and Planet Formation and Niels Bohr Institute, University of Copenhagen, Juliane Maries Vej 30, DK-2100, Copenhagen, Denmark; aake@nbi.dk}

\altaffiltext{4}{Theoretical Division, Los Alamos National Laboratory, Los Alamos, NM 87545, USA; dccollins@lanl.gov, hao\_xu@lanl.gov}

\altaffiltext{5}{ICREA \& ICC, University of Barcelona, Marti i Franqu\'{e}s 1, E-08028, Barcelona, Spain; ppadoan@icc.ub.edu}

\altaffiltext{6}{San Diego Supercomputer Center, 
University of California, San Diego; 10100 Hopkins Drive,  La Jolla, CA 92093-0505, USA; mlnorman@ucsd.edu}

\altaffiltext{7}{Kavli Institute for Particle Astrophysics and Cosmology, 
Stanford Linear Accelerator Center and Stanford Physics Department, Menlo Park, CA 94025, USA; tabel@stanford.edu}

\altaffiltext{8}{Zentrum f\"ur Astronomie der Universit\"at Heidelberg, Institut f\"ur Theoretische Astrophysik,
Albert-Ueberle-Str. 2, D-69120 Heidelberg, Germany; chfeder@ita.uni-heidelberg.de}

\altaffiltext{9}{Hamburger Sternwarte, Gojenbergsweg 112, D-21029 Hamburg, Germany; banerjee@hs.uni-hamburg.de}

\altaffiltext{10}{Max-Planck-Institute for Astronomy, K\"onigstuhl 17, D-69117 Heidelberg, Germany; flock@mpia.de}

\altaffiltext{11}{Ecole Normale Sup\'{e}rieure de Lyon, CRAL, 69364 Lyon, France}

\altaffiltext{12}{FLASH Center for Computational Science, 5747 S. Ellis Ave., Chicago IL
60637, USA; dongwook@flash.uchicago.edu}

\altaffiltext{13}{Astronomy Department, University of California, Berkeley, CA 94720, USA; psli@berkeley.edu}

\altaffiltext{14}{Max-Planck-Institut f\"ur Plasmaphysik, D-85748 Garching, Germany; Wolf.Mueller@ipp.mpg.de, cvogel@ipp.mpg.de}

\altaffiltext{15}{CEA, IRFU, SAp. F-91191 Gif-sur-Yvette, France; romain.teyssier@cea.fr}

\altaffiltext{16}{Institute of Theoretical Physics, University of Zurich, Winterthurerstrasse 190, 8057 Zurich, Switzerland}

\altaffiltext{17}{Keldysh Institute for Applied Mathematics, Russian Academy of Sciences, Miusskaya Pl. 4, Moscow 125047, Russia; ustyugs@keldysh.ru}

\begin{abstract}
Many astrophysical applications involve magnetized turbulent flows with shock waves. Ab initio
star formation simulations require a robust representation of supersonic turbulence in molecular
clouds on a wide range of scales imposing stringent demands on the quality of numerical algorithms.
We employ simulations of supersonic super-Alfv\'enic turbulence decay as a benchmark test problem
to assess and compare the performance of nine popular astrophysical MHD methods actively used to 
model star formation. The set of nine codes includes: ENZO, FLASH, KT-MHD, LL-MHD, PLUTO, PPML, 
RAMSES, STAGGER, and ZEUS. These applications employ a variety of numerical approaches, including 
both split and unsplit, finite difference and finite volume, divergence preserving and
divergence cleaning, a variety of Riemann solvers, a range of spatial reconstruction 
and time integration techniques. 
We present a comprehensive set of statistical measures designed to quantify the effects 
of numerical dissipation in these MHD solvers. We compare power spectra for basic fields 
to determine the effective spectral bandwidth of the methods and rank them based on their 
relative effective Reynolds numbers. 
We also compare numerical dissipation for solenoidal and dilatational velocity 
components to check for possible impacts of the numerics on small-scale density statistics. 
Finally, we discuss convergence of various characteristics for the turbulence decay test 
and impacts of various components of  numerical schemes on the accuracy
of solutions. The nine codes gave qualitatively the same results, implying 
that they are all performing reasonably well and are useful for scientific applications. 
We show that the best performing codes employ a consistently high order 
of accuracy for spatial reconstruction of the evolved fields, transverse gradient 
interpolation, conservation law update step, and Lorentz force computation.
The best results are achieved with divergence-free evolution of the magnetic field 
using the constrained transport method, and using little to no explicit artificial 
viscosity. Codes which fall short in one or more of these areas are still useful, but 
they must compensate higher numerical dissipation with higher numerical resolution. 
This paper is the largest, most comprehensive MHD code comparison on an application-like 
test problem to date. We hope this work will help developers improve their 
numerical algorithms while helping users to make informed choices in picking optimal 
applications for their specific astrophysical problems.

\end{abstract}


\section{Introduction}
It is well established that the observed supersonic turbulence plays an important role in the fragmentation
of molecular clouds leading to star formation \citep{maclow.04,mckee.07}. As illustrated by numerical simulations, random
supersonic flows in an isothermal gas result in a complex network of shocks creating a filamentary density structure
with a very large density contrast \citep[e.g.,][see also references in \citet{klessen..09,pudritz10}]{kritsuk...07,federrath..08}. 
Because it can naturally generate density
enhancements of sufficient amplitude to allow the formation of low mass stars or even brown dwarfs within complex layers of
post-shock gas, the turbulence may directly affect the mass distribution of pre-stellar cores and stars
\citep{padoan.02,padoan07,hennebelle.08,hennebelle.09}. Furthermore, the turbulence must be at least partly
responsible for the low star formation rate per free-fall time observed in most environments \citep{krumholz.07},
because the turbulent energy generally exceeds the gravitational energy on small scales within molecular clouds
(the virial parameter is almost always larger than unity, as shown by \citet{falgarone..92} and 
\citet{rosolowsky...08} in Perseus).
Theoretical models of the star formation rate based on the effect of turbulence have recently been proposed
\citep{krumholz.05,padoan.09}.

The importance of turbulence in the process of star formation provides an opportunity for theoretical modeling,
because one can assume that molecular clouds follow the universal statistics of turbulent flows, for example with
respect to the probability density function (PDF) of gas density and the scaling of velocity differences. 
The turbulence is also a challenge for numerical
simulations of star formation, because the limited dynamical range of the simulations cannot always approximate
well enough the scale-free behavior of the turbulent flow. 
The Kolmogorov dissipation scale, $\eta_{\rm K}$, is the smallest turbulent scale below which viscous dissipation 
becomes dominant. It can be computed as $\eta_{\rm K} = (\nu^3/\epsilon)^{1/4}$, where $\nu$ is the 
kinematic viscosity and $\epsilon$ the mean dissipation rate of the turbulence. 
The kinematic viscosity can be approximated as $\nu\approx \upsilon_{\rm th}/(\sigma n)$,
where $\upsilon_{\rm th}$ is the gas thermal velocity, $n$ is the gas mean number density, 
and $\sigma\approx 5\times 10^{-15}$~cm$^2$ is the gas collisional cross section. 
The mean dissipation rate can be estimated as $\epsilon\sim \upsilon^3/\ell$, where $\ell$ is a 
scale within the inertial range of the turbulence, and $\upsilon$ is the rms velocity at the scale $\ell$.
In molecular clouds, assuming the \citet{larson81} relations $\upsilon\sim 1$~km~s$^{-1}(\ell/1$pc$)^{0.42}$ and $n \sim 10^3$~cm$^{-3}(\ell/1$pc$)^{-1}$, 
a gas temperature of 10~K, and a driving scale of $\sim70$~pc, we obtain $\eta_{\rm K}\sim 10^{14}$~cm, well below 
the characteristic spatial resolution of the gas dynamics in star formation simulations. 

The dynamic range limitation of the simulations can be expressed in terms of the Reynolds number. The Reynolds
number estimates the relative importance of the nonlinear advection term and the viscosity term in the
Navier--Stokes equation, $\mathrm{Re}=\upsilon_{\rm rms}{\cal L}/\nu$, where $\upsilon_{\rm rms}\equiv\sqrt{\left<\upsilon^2\right>}$ 
is the flow rms velocity, ${\cal L}$ is the integral scale of the turbulence (of the order of the energy injection 
scale). The Reynolds number can also be expressed as $\mathrm{Re}=({\cal L}/\eta_{\rm K})^{4/3}$. Based on the same 
assumptions used above to derive $\eta_{\rm K}$, we obtain $\mathrm{Re}\sim10^8$ for typical molecular cloud values. 
At present, the largest simulations of supersonic turbulence may achieve an effective Reynolds
number $\mathrm{Re}\sim 10^4$ \citep{kritsuk...09a,jones...11}.

Numerical simulations are incapable of describing the smallest structures of magnetic fields in star-forming clouds.
The characteristic magnetic diffusivity, $\eta$, of the cold interstellar gas is much smaller than the kinematic
viscosity, $\nu$. As a result, magnetic fields can develop complex structures on scales much smaller than the
Kolmogorov dissipation scale, $\eta_{\rm K}$, where the velocity field is smooth. Introducing the magnetic Prandtl
number, $\mathrm{Pm}$, defined as the ratio of viscosity and diffusivity, $\mathrm{Pm}=\nu/\eta$, this regime is
characterized by the condition $\mathrm{Pm}\gg1$. The magnetic diffusivity can be expressed as 
$\eta = c^2 m_{\rm e} \nu_{\rm en} / 4 \pi n_{\rm e} e^2$ (cgs), where $c$ is the speed of light, $m_{\rm e}$ 
the electron mass, $\nu_{\rm en}$ the collision frequency of electrons with neutrals, $n_{\rm e}$ the number 
density of electrons, and $e$ the electron charge. This expression neglects electron-ion collisions, because at 
the low ionization fractions and temperatures of molecular clouds the dominant friction force on the electrons 
is from collisions with neutrals. The collision frequency of electrons with neutrals can be written as
$\nu_{\rm en} = n_{\rm n} \sigma \upsilon_{\rm th,e}$, where $n_{\rm n}$ is the number density of neutrals 
($\sim n$ in molecular clouds), $\sigma$ is the gas collision cross section given above, and $\upsilon_{\rm th,e}$ 
is the thermal velocity of the electrons.  The magnetic Prandtl number is then given by
$\mathrm{Pm} \approx  2 \times 10^5 (x_{\rm i}/10^{-7}) (n/1000$cm$^{-3})^{-1}$, where $x_{\rm i}$ is 
the ionization fraction.

Numerical simulations without explicit viscosity and magnetic diffusivity usually have effective values of $\mathrm{Pm}\sim1$,
very far from the conditions in molecular clouds. If the magnetic field strength is determined self-consistently
by a small-scale turbulent dynamo, this numerical limitation may cause an artificially low magnetic field strength
in low-resolution simulations, or in simulations based on MHD solvers with large effective magnetic diffusivity.
Such simulations may not reach the critical value of the magnetic Reynolds number, $\mathrm{Rm}$, required by the
turbulent dynamo. The magnetic Reynolds number is defined as $\mathrm{Rm=Re Pm}= \upsilon_{\rm rms}{\cal L}/\eta$. Its critical
value for the turbulent dynamo in supersonic turbulence was found to be $\mathrm{Rm}_{\rm crit}\approx 80$ in the regime
with $\mathrm{Pm}\sim 1$ and for a sonic rms Mach number $M_{\rm s}\approx 2.5$,
where $M_{\rm s}=\upsilon_{\rm rms}/c_{\rm s}$ is the ratio of the flow rms velocity 
and the speed of sound \citep{haugen..04}. \citet{federrath....11} find $\mathrm{Rm}_{\rm crit}\approx 40$ 
for transonic turbulence, driven by the gravitational collapse of a dense, magnetized gas cloud.

Besides the effective $\mathrm{Re}$ and $\mathrm{Pm}$, the other two non-dimensional parameters of isothermal MHD turbulent
simulations are the rms sonic Mach number, defined above, and the rms Alfv\'{e}nic Mach number, $M_{0,\rm A}=\upsilon_{\rm rms}/\upsilon_{0,\rm A}$,
where $\upsilon_{0,\rm A}$ is the mean Alfv\'{e}n speed defined as $\upsilon_{0,\rm A}=B_0/\sqrt{4\pi \rho_0}$, and $B_0$ and $\rho_0$
are the mean magnetic field and mean gas density, respectively. The initial conditions of the numerical test described
in this work have $M_{\rm s}\approx 9$ and $M_{0,\rm A}\approx 30$. In the test runs, the value of $M_{\rm s}$
decreases with time (no driving force is used), as shown in the left panels of Figure~\ref{energy}. $\upsilon_{0,\rm A}$ is instead
constant, because both $B_0$ and $\rho_0$ are conserved quantities in the simulations. However, the rms value of
the magnetic field strength, $\sqrt{\left< B^2 \right>}$, depends on both $B_0$ and $\upsilon_{\rm rms}$. In these simulations
$B_0$ is very low, and the turbulence is highly super-Alfv\'{e}nic, meaning that $\upsilon_{\rm rms} \gg \upsilon_{0,\rm A}$. In this regime,
the magnetic field is locally amplified by compression and stretching resulting in a statistically steady state
with $\sqrt{\left< B^2 \right>}\gg B_0$. 
The rms Alfv\'{e}nic Mach number defined %instead 
in terms of the mean magnetic and kinetic energies, $M_{\rm A}=\sqrt{\left<\rho \upsilon^2\right>/\left<B^2/4\pi\right>}\approx 4.4$ and decreases 
with time as the turbulence decays, as shown by the right panels of Figure~\ref{energy}.

Based on the observed dependence of the velocity dispersion on spatial scale in molecular clouds
\citep[e.g.,][]{larson81,heyer.04}, the initial value of $M_{\rm s}$ in our test runs is relevant to 
star-forming regions on scales of a few parsecs. The super-Alfv\'{e}nic nature of molecular cloud turbulence was suggested
by \citet{padoan.99}, and has received further support in more recent work \citep{lunttila...08,lunttila...09,padoan......10,kritsuk..11}.

One way to assess the ability of numerical simulations to approximate the behavior of turbulent flows
is to study the power spectra of relevant quantities, such as velocity and magnetic fields. The interpretation
of velocity power
spectra from numerical simulations face the following problems: (1) the limited extent of the inertial range
of turbulence due to the limited range of spatial scales discussed above (or even the complete absence of an
inertial range in the case of low resolution simulations); (2) the emergence of the bottleneck effect in
hydrodynamic simulations \citep[e.g.,][]{falkovich94,dobler...03,haugen.04} as soon as the numerical resolution
is large enough to generate an inertial range; (3) the dependence of the power spectrum on the numerical
schemes; (4) the dependence of the numerical resolution necessary for convergence
on the numerical method.

This work addresses the above problems and the general issue of the quality of MHD codes with respect to the
description of highly supersonic and super-Alfv\'{e}nic isothermal turbulent flows. 
We do not study the quality of simulations of the gravitational 
collapse of gravitationally unstable regions with adaptive mesh 
refinement or Lagrangian methods in this paper. Although most star formation 
simulations eventually take advantage of such techniques, here 
we focus on the simulations of turbulent 
flows where gravity is neglected. This work considers
high-resolution simulations of MHD turbulence, while related studies of nonmagnetized flows have been 
recently published by \cite{kitsionas+12-08} and \cite{price.10}.

The paper is organized as follows. In Section~2, we describe the simulation setup.
In Section~3, we introduce the algorithms used.  In Section~4, we discuss
the diagnostic techniques utilized in the paper.  In Section~5 we
present the results from each code, and in Section~6 we discuss the
impact of method design on the numerical dissipation properties.  
Finally, Section~7 summarizes our conclusions.

\section{The Turbulence Decay Test Problem}
Modern numerical methods for astrophysical turbulence simulations are designed to produce approximations to the
limit of viscous and resistive solutions as the viscosity and magnetic diffusivity are reduced to zero.
Numerical experiments carried out with such methods can be viewed as implicit large eddy simulations, or  
ILES \citep{grinstein..07}. \citet{sytine....00} demonstrated that Euler solvers, like PPM \citep{colella.84}, 
are more efficient than Navier--Stokes solvers in providing a better scale separation at a given grid resolution
\citep[see also][]{benzi.....08}. 
Here we employ the same ILES technique for MHD simulations of decaying supersonic turbulence. The numerical methods we 
compare differ in their implicit subgrid models and the focus of 
this paper is on understanding the origin of those differences, which could help to improve our methods.

We, thus, solve numerically the system of MHD equations for an ideal isothermal gas in a cubic 
domain of size $L$ with periodic boundary conditions:
\begin{equation}
\frac{\partial \rho}{\partial t}+{\bf \nabla}\cdot(\rho {\bf u}) =0,
\end{equation}
\begin{equation}
\frac{\partial \rho {\bf u}}{\partial t}+{\bf \nabla\cdot}\left[\rho {\bf uu} -
{\bf BB}+ \left(p+\frac{{\bf B}^2}{2}\right){\bf I}\right]=\rho{\bf F},\label{mome}
\end{equation}
\begin{equation}
\frac{\partial {\bf B}}{\partial t}+{\bf \nabla}\cdot({\bf uB} - {\bf Bu}) = {\bf 0}.\label{fara}
\end{equation}
Here, $\rho$ and ${\bf u}$ are the gas density and velocity, ${\bf B}$ is the magnetic field strength, 
$p$ is the gas pressure, and ${\bf I}$ is the unit tensor. 

All numerical methods discussed in this paper are designed to conserve mass, momentum 
and magnetic flux, and attempt to keep ${\bf \nabla\cdot B}=0$ to the machine precision. 
All methods are formulated to approximate the ideal MHD Equations (1)--(3). However, due to 
the finite numerical viscosity and magnetic diffusivity, as well as artificial viscosity and diffusivity 
added for numerical concerns, the actual equations evolved will have additional dissipation terms 
on (2) and (3). The exact nature of these dissipation terms is method-dependent.

In this section and below, we use dimensionless code units, such that the domain 
size $L=1$; the gas density $\rho$ is given in units of the mean gas density $\rho_0$; the
gas pressure $p$ is given in the units of the uniform initial pressure $p_0$, and the velocity 
${\bf u}$ is given in units of the sound speed, $u=\upsilon/c_{\rm s}$. The uniform mean magnetic field 
is $B_0=\sqrt{2/\beta_0}=0.3$, where the ratio of thermal-to-magnetic pressure $\beta_0=22$. The code
units also imply that ${\bf B}$ incorporates the $1/4\pi$ factor so that the magnetic pressure
is given by $B^2/2$ in the code units.

Initial conditions for the decay test were generated in 2007 with an earlier, non-conservative version
of the STAGGER code on a $1000^3$ grid using a time-dependent random large-scale 
($k/k_{\rm min}\le2$, where $k_{\rm min}=2\pi/L$) isotropic solenoidal force (acceleration) 
${\bf F}$ to stir the gas and reach an rms sonic Mach number $M_{\rm s,0}\approx9$. There was no forcing
in the induction Equation (3), so the rms magnetic field was passively amplified through interaction
with the velocity field.
The model was initiated with a uniform density $\rho_0$ and pressure $p_0$, random large-scale velocity field 
${\bf u}_0$, and a uniform magnetic field ${\bf B}_0$ aligned with the $z$-coordinate direction. 
To achieve a saturated turbulent state, the flow was evolved with the STAGGER code for three dynamical times 
(defined as $t_{\rm d}\equiv L/2M_{\rm s,0}$). Assuming an initial $M_{\rm s,0}=10$, $t_{\rm d}=0.05$ in the code units determined 
by the box sound crossing time. In the saturated turbulent state, the level of magnetic fluctuations is
$\sim50$ times higher than $B_0$, i.e., ${\bf B} = {\bf B}_0 + {\bf b}$, where $b_{\rm rms}\gg B_0$ and $\left<{\bf b}\right>\equiv{\bf 0}$.

The actual test runs were performed at grid resolutions of $256^3$, $512^3$, and in a few cases (PPML and ZEUS) 
also $1024^3$ cells. Data regridding utilized conservative interpolation of hydrodynamic variables while a vanishing
${\bf \nabla\cdot B}$ in the interpolated initial states was enforced with ${\bf \nabla\cdot B}$ cleaning.
The evolution of decaying turbulence (${\bf F}\equiv{\bf 0}$) was followed for $\Delta t=0.2=4t_{\rm d}$ and 10 flow 
snapshots equally spaced in time were recorded for subsequent analysis. The timing of these snapshots in the adopted
time units is as follows: $t_1=0.02$, $t_2=0.04$, \ldots, $t_{10}=0.2$, assuming $t=0$ corresponds to the end of the
initial forcing period.

\section{Numerical Methods and Implementations}
\label{sec.methods}

\subsection{ENZO 2.0}
ENZO's \citep{osheaetal05} MHD scheme \citep{wang.09} employs the following
components: second-order spatial interpolation via the Piecewise
Linear Method \citep{vanleer79}; second-order time integration via a
second-order Runge--Kutta method \citep{shu.88}; the HLL Riemann solver for
computation of interface fluxes \citep{harten..83}; and the \citet{dedner.....02}
scheme for maintaining the divergence of the magnetic field close to zero.
The code is formally second-order accurate in
both time and space. These one-dimensional components are combined to
form a three-dimensional method in a directionally unsplit manner, with the 
Runge--Kutta integration mediating the wave information between the three
flux computations.  The slope limiter $\theta$, which controls the
sharpness of the reconstruction, was set at 1.5 as in \citep{wang.08}.
Larger values were tried for $256^3$ grids, without significant change 
to the solution.

\subsection{FLASH 3}
The FLASH3 \citep{FryxellEtAl2000,DubeyEtAl2008} simulations presented in 
this study have used a completely new MHD scheme implementation \citep{LeeDeane2008}. 
The solver adopts a dimensionally unsplit integration on a staggered grid 
(Unsplit Staggered Mesh), for the multidimensional MHD formulation, based on 
a finite-volume, higher-order Godunov method. A new second-order data 
reconstruction-evolution method, extended from the corner transport upwind 
(CTU) approach of \citet{Colella90} has been used, which guarantees proper evolution of 
in-plane dynamics of magnetic fields. The importance of the in-plane
field evolution is described and tested in the field-loop advection test in
\citet{2005JCoPh.205..509G}. The unsplit staggered mesh solver (USM) has also shown a successful
performance on this test, maintaining a correct in-plane field dynamics 
\citep{LeeDeane2008}. The algorithm uses a new ``multidimensional 
characteristics analysis'' in calculating transverse fluxes. This approach is 
advantageous and efficient because it does not require solving a set of Riemann 
problems for updating transverse fluxes. High Mach number turbulent flows 
require a precise and positive-preserving solver capable of resolving complex 
shock structures while keeping numerical diffusion as small as possible. 
We therefore chose the HLLD Riemann solver \citep{MiyoshiKusano2005}, which 
greatly improves the robustness and accuracy of supersonic MHD turbulence 
simulations as the Roe solver easily fails to preserve positive states of 
density and/or pressure in strong rarefaction waves. For further enhancing 
solution accuracy and stability, we chose a hybrid limiter that uses the 
compressive van Leer's slope limiter for linearly degenerate waves and the 
more diffusive minmod limiter for genuinely nonlinear waves.

\subsection{KT-MHD}
The KT-MHD code is an implementation of a semidiscrete central-difference 
scheme developed by \citet{2000JCoPh.160..241K}. The total time derivative of 
the hydrodynamic quantities is computed using the flux definition of the 
Kurganov--Tadmor scheme, a higher order extension of the Lax--Friedrichs scheme. 
The flux values are evaluated at the cell interfaces. The corresponding point 
values of the conserved quantities are interpolated to the cell interfaces 
via a third-order CWENO scheme in three space dimensions following \citet{1132394}. 
The averages of the magnetic field components reside at the cell interfaces 
and are reconstructed in diagonal direction, also using a third-order CWENO scheme. 
The smoothness indicators (and thereby the nonlinear weights) of the CWENO scheme 
are based on the density field, only. Componentswise smoothness indicators have 
shown to lead to a much higher numerical viscosity. The total time derivative of 
the magnetic field is computed by a constrained transport (CT) method of 
\citet{1007844}. The resulting set of ordinary differential equations is 
integrated in time by a fourth-order Runge--Kutta scheme. 
The code uses a regular grid and the so-called {\em pencil} decomposition 
in its MPI-parallel implementation.
The idea of combining the Kurganov--Tadmor central-difference scheme with a 
CT method for the magnetic field update was first implemented 
by Ralf Kissmann and published in his PhD thesis \citep{RKPHD} and is used by 
\citet{1141227} in their Racoon code. 
 
\subsection{LL-MHD}
The CT-based LL-MHD solver \citep{collins....09} employs 
a divergence preserving higher-order Godunov method 
of \citet{2008ApJS..174....1L}, which uses second-order spatial
reconstruction and second order time reconstruction to compute the
interface states, and the isothermal HLLD Riemann solver of
\citet{2007JCoPh.225.1427M} to compute the flux from those
reconstructed states. This is done in a directionally split fashion,
with the order permutation of Strang to preserve the second-order accuracy.  
The solver uses the CT method of \citet{2005JCoPh.205..509G} 
to maintain the divergence-free evolution of the magnetic field. 
LL-MHD is also installed in the AMR code ENZO, and has been used to
study a range of astrophysical phenomena, from galaxy clusters \citep{xu10} 
to pre-stellar cores \citep{collins10}.

\subsection{PLUTO 3.1}
The PLUTO code \citep{mig09} is a highly modular, multi-dimensional and multi-geometry
code that can be applied to relativistic or non-relativistic MHD or HD (hydrodynamic)
flows. PLUTO comprises several numerical methods, like the high-order conservative 
finite-difference divergence cleaning MHD method \citep{mig10I} as well as 
finite-volume CTU schemes \citep{mig10II}. 
The latest version of the PLUTO code (V. 3.1---2010 August) allows to choose 
between several space reconstruction and time integration methods as well as 
several approximate Riemann solvers including HLL, HLLC, HLLD or the Roe Riemann solver.
For the MHD formulation one can choose between the eight-wave formulation 
\citep{Powell99}, the divergence cleaning method \citep{dedner.....02}, and the 
CT method.
The possibility to switch between several numerical methods allows to handle a
wide range of astrophysical problems. 
For this test we used the accurate Roe Riemann solver in combination
with a third-order reconstruction \citep{cad09}, characteristic variable
limiting, the Runge--Kutta 3 time integration and the \citet{Powell99} eight-wave MHD
formalism; three-dimensional effects were incorporated by way of the 
Runge--Kutta integration, without the use of the transverse flux gradients used in CTU.
Courant number was set to 0.3. 

\subsection{PPML}
The piecewise parabolic method on a local stencil \citep[PPML,][]{ustyugov...09}
is a compact stencil variant of the popular PPM algorithm \citep{colella.84} for compressible 
magnetohydrodynamics.  The principal difference between PPML and PPM is 
that cell interface states are evolved rather that reconstructed at every time step, resulting 
in a more compact stencil. The interface states are evolved using Riemann invariants containing all 
transverse derivative information. The conservation laws are updated in an unsplit fashion, 
making the scheme fully multidimensional. Divergence-free evolution of the magnetic field is 
maintained using the higher-order-accurate CT technique of 
\citet{2005JCoPh.205..509G}. The method employs monotonicity constraints to preserve the order 
of scheme in points of local extrema \citep{suresh.97,balsara.00,rider..07}. 
To preserve monotonicity in multidimensions a method from \citet{barth90} is additionally applied.
An updated component of the electric field at a cell boundary is 
calculated by averaging the quantities obtained from known components 
of flux-vectors and values of gradient of the electric field \citep{2005JCoPh.205..509G}. 
The performance of PPML was tested on several numerical problems, which demonstrated
its high accuracy on both smooth and discontinuous solutions \citep{ustyugov...09}.
Simulations of supersonic magnetized turbulence in three dimensions with PPML 
show that low dissipation and wide spectral bandwidth of this method make it an ideal 
candidate for direct turbulence simulations \citep{kritsuk...09a,kritsuk...09b}.

\subsection{RAMSES}
RAMSES \citep{Teyssier2002} is an unsplit Godunov AMR scheme with a second-order total variation
diminishing spatial reconstruction using the Monotonized Central slope limiter. 
Magnetic field is updated using the
CT method, using two-dimensional Riemann problem at cell edges to compute
the electro-motive-force that enters into the induction equation. The magnetic field divergence, expressed
in an integral form on cell faces, is therefore zero down to machine accuracy. Conservative variables
are updated by solving one-dimensional Riemann problems at cell faces. Both the one-dimensional and the two-dimensional Riemann solvers
are based on the HLLD MHD approximate Riemann solution \citep{MiyoshiKusano2005}. 
More details on the MHD scheme can be found
in Teyssier et al. (2006) and Fromang et al. (2006). 

\subsection{STAGGER}
The STAGGER Code is originally based on a code developed as part of the
PhD thesis of Klaus Galsgaard \citep{galsgaard96}.  Several versions exist,
and the code is used in many different circumstances 
\citep{galsgaard.96,Padoan97,Padoan98,Padoan00,Stein98,%
asplund...00,padoan...04,gudiksen.05,braithwaite.06,archontis..07,lunttila...09,stein...11,padoan.09}.

In the context of solar and stellar physics it is equipped with a 
multi-frequency radiative transfer module and a comprehensive equation 
of state module that includes a large number of atomic and molecular species, 
to be able to compute realistic three-dimensional models of the near-surface layers of stars.
The widths, shifts, and asymmetries of synthetic spectral lines computed 
from such models exemplifies some of the most precise agreements between 
three-dimensional numerical simulations and astrophysical observations \citep{asplund...00}.

In the context of supersonic turbulence studies earlier works 
\citep{Padoan97,Padoan98,Padoan00} were based on a non-conservative version of 
the code, which evolved the primitive variables $\lnr$, $\uu$, and $\BB$.  
The ``per-unit-mass''  formulation based on these variables is simple and robust, 
but has the disadvantage that mass and momentum are not conserved exactly by the 
discretized equations.

The current version of the code instead uses the per-volume variables $\rho$, 
$\rho \uu$, and $\rho E$, where $E$ is the internal energy per unit mass, allowing
a discretization that explicitly conserves mass, momentum, energy, and magnetic flux.  
In the isothermal case of relevance here the code solves these equations:
\eab{1}
\ddt{\rho} & = & - \div \rho \uu, 
\\
\ddt{\rho \uu} & = & - \div (\rho \uu \uu + \TT ) - \grad p + (\curl \BB) \times \BB,
\\
\ddt{\BB} & = & - \curl ( -\uu \times \BB + \eta \curl \BB ) ,
\ean
where $\TT$ is the viscous stress tensor, which we write as
\eb{2}
\Tij = - \rho \nu \Sij ,
\en
and $\Sij$ is the strain rate
\eb{3}
\Sij = \half \left( \ddi{u_j}+\ddj{u_i} \right).
\en
The viscosity $\nu$ and magnetic diffusivity $\eta$ are spatially dependent, in a 
manner reminiscent of the Richtmyer \& Morton formulation, with
\eb{4}
%\nu = ( n_1 \uw + n_2 \uc + n_S \uS ) \ds
\nu = ( n_1 \uw + n_2 \uc ) \ds ,
\en
where $\ds$ is the mesh size, $\uw$ is the wave speed, $\uc$ is the positive part of a second order 
approximation of $-\ds \div\uu$.
The magnetic diffusivity is taken to be
\eb{5}
\eta = n_B ( n_1 \uw + n_2 \delta u_B^+ ) \ds ,
\en
where $\delta u_B^+$ is analogous to $\uc$, except only the component of the velocity
perpendicular to $\BB$ is counted.

Here, $n_1$, $n_2$, and $n_B$ are numerical coefficients of the order of unity. The $n_1\uw$
term, where $n_1\sim 0.03$ is a relatively small constant, is 
needed to provide stabilization and a weak dispersion of linear waves, while
the $n_2 \uc$ term, with $n_2\sim 0.5$, provides enhanced dissipation 
in shocks, where the rate of convergence $-\div\uu$ is large. 
Since the magnetic field is insensitive to motions parallel to the field,
only perpendicular motions are gauged by the corresponding magnetic diffusivity term.
$n_B$ is essentially an inverse magnetic Prandtl number. 

The tensor formulation of the viscosity ensures that the viscous force is 
insensitive to the coordinate system orientation, thereby avoiding artificial 
grid-alignment.

\subsection{ZEUS-MP}
ZEUS-MP is a widely used, multiphysics, massively parallel, message-passing 
Eulerian code for astrophysical fluid dynamic simulations in three dimensions. 
ZEUS-MP is a distributed memory version of the shared-memory code ZEUS-3D that 
uses block domain decomposition to achieve scalable parallelism. The code 
includes hydrodynamics, magnetohydrodynamics, and 
self-gravity. The HD and MHD algorithms are based on the method of finite 
differences on a staggered mesh \citep{stone.92a,stone.92b}, which incorporates 
a second-order-accurate, monotonic advection scheme \citep{1977JCoPh..23..276V}. The 
MHD algorithm is suited for multidimensional flows using the method 
of characteristics scheme (MOC-CT) first suggested by \citet{1995CoPhC..89..127H}. Advection is 
performed in a series of directional sweeps that are cyclically permuted at 
each time step. Because ZEUS-MP is designed for large simulations on parallel 
computing platforms, considerable attention is paid to the parallel performance 
characteristics of each module in the code. Complete discussion on all algorithms 
in ZEUS-MP can be found in \cite{2006ApJS..165..188H}. All the MHD turbulence decay 
simulations performed using ZEUS-MP in this paper use a quadratic 
(von Neumann-Richtmyer) artificial viscosity coefficient qcon of 2.0 and a 
Courant number of 0.5.

\ctable[star,pos=t, caption = Selected Numeric Values for the Decay Test]{l c c c c c c c}{
\tnote[{\rm a}]{Mean specific kinetic energy density at $t=0.2$ normalized by the reference solution; see Section~5.1 and Figure~1.}
\tnote[{\rm b}]{Mean magnetic energy density at $t=0.2$ normalized by the reference solution; see Section~5.1 and Figure~1.}
\tnote[{\rm c}]{A proxy for the mean dissipation rate of specific kinetic energy at $t=0.02$; see Section~5.2 and Figure~2 left.}
\tnote[{\rm d}]{A proxy for the mean dissipation rate of magnetic energy at $t=0.02$; see Section~5.2 and Figure~2 right.}
\tnote[{\rm e}]{Effective spectral bandwidth for the velocity; see Section~5.4 and Figure~4 left. }
\tnote[{\rm f}]{Effective spectral bandwidth for the magnetic field; see Section~5.4 and Figure~4 right. }
\tnote[{\rm g}]{Ratio of dilatational-to-solenoidal power averaged over $k/k_{\rm min}>100$ at $t=0.2$; see Section~5.5 and Figure~5 right.}
}{
\hline
\hline
Code                                      & 
$E_{\rm K}/E_{\rm K, ref}$ \tmark[{\rm a}]  & 
$E_{\rm M}/E_{\rm M, ref}$ \tmark[{\rm b}]  & 
$2\Omega+4/3\Delta$ \tmark[{\rm c}]             & 
$J^2$ \tmark[{\rm d}]                           & 
${\bf u}$-bandwidth \tmark[{\rm e}]             & 
${\bf B}$-bandwidth \tmark[{\rm f}]             &
%$k_{25}({\bf u})/k_{\rm N}$ \tmark[{\rm e}]      & 
%$k_{25}({\bf B})/k_{\rm N}$ \tmark[{\rm f}]      &
$\bar{\chi}(k>100k_{\rm min})$ \tmark[{\rm g}] \\
\hline
\ENZO    & 1.001 & 0.78 & 0.93 & 0.92 & 0.19 & 0.07 & 0.60 \\ % 0.49
\FLASH   & 1.000 & 0.94 & 0.85 & 1.38 & 0.15 & 0.20 & 0.27 \\ % 0.50
\KTMHD   & 1.041 & 0.85 & 0.89 & 1.30 & 0.20 & 0.13 & 0.86 \\ % 0.52
\LLMHD   & 1.062 & 0.81 & 1.02 & 0.80 & 0.22 & 0.10 & 0.29 \\ % 0.44
\PLUTO   & 1.077 & 0.92 & 1.03 & 1.14 & 0.20 & 0.12 & 0.32 \\ % 0.40
\PPML    & 1.043 & 0.92 & 1.20 & 1.46 & 0.24 & 0.20 & 0.32 \\ % 0.44
\RAMSES  & 1.069 & 0.87 & 1.07 & 1.18 & 0.24 & 0.09 & 0.33 \\ % 0.41
\STAGGER & 1.005 & 0.70 & 1.93 & 0.79 & 0.28 & 0.07 & 0.31 \\ % 0.35
\ZEUS    & 1.037 & 0.83 & 0.76 & 1.01 & 0.16 & 0.10 & 0.27 \\ % 0.39
\hline
}

\section{Data analysis}
\subsection{Power Spectra}
Given a vector field $\uu(\rr)$ discretized on a mesh $\ijk$ with $\uu_\ijk$
one can compute a power spectrum from the three-dimensional Fourier transform
$\ut_\ijk$ by summing the magnitudes squared, $|\ut_\ijk|^2$, over k-shells with
$K_n \leq |\kk| < K_{n+1}$.   If the Fourier transform coefficients $\ut_\ijk$
are normalized so the rms value of the corresponding function in real space
is equal to unity, then the sum of the squares in Fourier space is equal to the
average of the function squared in real space (Parseval's relation):
\eb{p1}
\RMS^2 = \sum_\ijk |\ut_\ijk|^2 = \frac{1}{N} \sum_\ijk |\uu_\ijk|^2 ,
\en
where $N$ is the total number of $\ijk$ points.  

In the codes used to
analyze the results for the current paper we use the real valued Fast 
Fourier Transform routine {\tt srfftf} from the {\tt fftpack} software 
package, which returns coefficients $a_k=N/2$ for sine and cosine functions, 
except for the DC and Nyquist components, which return coefficients $N$.
Proper power normalization requires that sine and cosine components contribute
power $1/2$, and the returned coefficients should thus be multiplied with 
$\sqrt{2}/N$, except for the DC and Nyquist components (which are the first 
and last coefficients returned from {\tt srfftf}), which should be multiplied
with $1/N$.

The power spectrum $P(k)$ expresses how much of the power falls in each $k$-interval.  
If the power is collected in discrete bins,
\eb{p2a}
P_n = \sum_{K_n \leq |\kk| < K_{n+1}} |\ut_\ijk|^2 ,
\en
then the total power can also be expressed as
\eb{p2}
\RMS^2 = \sum_n P_n ,
\en
where the sum is taken over all bins.

To illustrate the power spectrum $P_n$ graphically one needs to assign 
a wavenumber $k_n$ to each bin.  A natural but not quite optimal choice
is to use the midpoint of the bin; $k_n=(K_n+K_{n+1})/2$.   A better choice
is to use the mean of the wavenumbers that actually fall inside the bin 
(to see why this is better consider a case with very wide bins and a
function with power at only a few discrete wavenumbers).

It turns out that one gets smoother power spectra if one assigns a value
\eb{p3}
P_n' = \frac{4\pi}{3} (K_{n+1}^3-K_n^3) \frac{1}{N_\bin} \sum_\bin |\ut_\ijk|^2
\en
rather than 
\eb{p4}
P_n = \sum_\bin |\ut_\ijk|^2
\en
to each bin.  In other words:  power spectra (at least those measuring 
fluid flow properties) become smoother if they measure the {\em average 
power} in a shell (times the shell volume) rather than the total power.
One can interpret this to mean that fluid flow properties are encoded 
in Fourier amplitudes as a function of wavenumber, rather than in total
power of Fourier amplitudes in a shell.  If (and only if) this is the 
case then the power spectrum fluctuates (as observed) down (or up) if 
by chance a shell contains fewer (or more) discrete wavenumbers than expected.

To be able to recover $P_n$ from $P_n'$ (e.g. for use in Parseval's relation) 
it is necessary to record the number of discrete Fourier amplitudes in 
each bin, $N_\bin$ in Equation~(\ref{eq:p3}) above.

Note also, that in order for Parseval's relation to be exact for three-dimensional power 
spectra, all Fourier components need to be included, which means that the
$k$-scale should really extend to a maximum value of $\sqrt{3} \, k_{\rm N}$, 
where $k_{\rm N}$ is the one-dimensional Nyquist frequency.  Nevertheless,
here we follow the common practice to truncate the three-dimensional power spectra at 
$k_{\rm N}$.

\subsection{Helmholtz Projections}
The divergence of a vector field $\ff_\ijk$, with Fourier transform coefficients $\ft_\ijk$, is
\eb{1}
\FT(\div\ff) = \ik \cdot \ft_{\ijk}.
\en
The vector coefficients $\ft_\ijk$ may be split into a component parallel 
to $\kk$ and a remaining component, which is perpendicular to $\kk$: 
\eb{2}
\ft_\ijk^{\parallel} = \kk (\kk\cdot\ft_\ijk) / |\kk|^2 ,
\en
and
\eb{3}
\ft_\ijk^{\perp} = \ft_\ijk - \ft_\ijk^{\parallel} .
\en
Taking the divergence of the latter, we have
\eb{4}
\ik\cdot\ft_\ijk^{\perp} = \ik\cdot\ft_\ijk - \ik\cdot\kk (\kk\cdot\ft_\ijk) / |\kk|^2 = 0 .
\en
The inverse transform based on the $\ft_\ijk^{\perp}$ coefficients is thus
solenoidal, while the inverse transform based on $\ft_\ijk^{\parallel}$ is purely
compressional.

\begin{figure*}
\epsscale{1.2}
\centerline{\plottwo{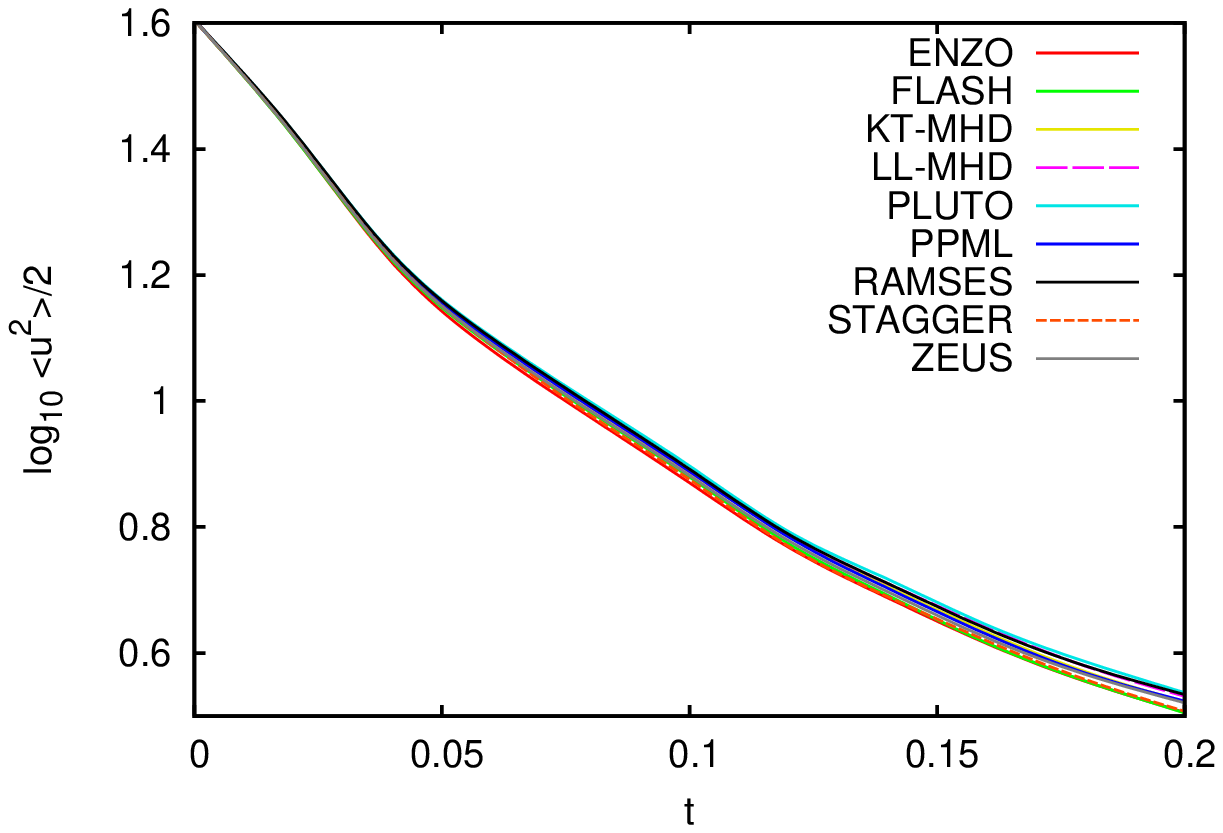}{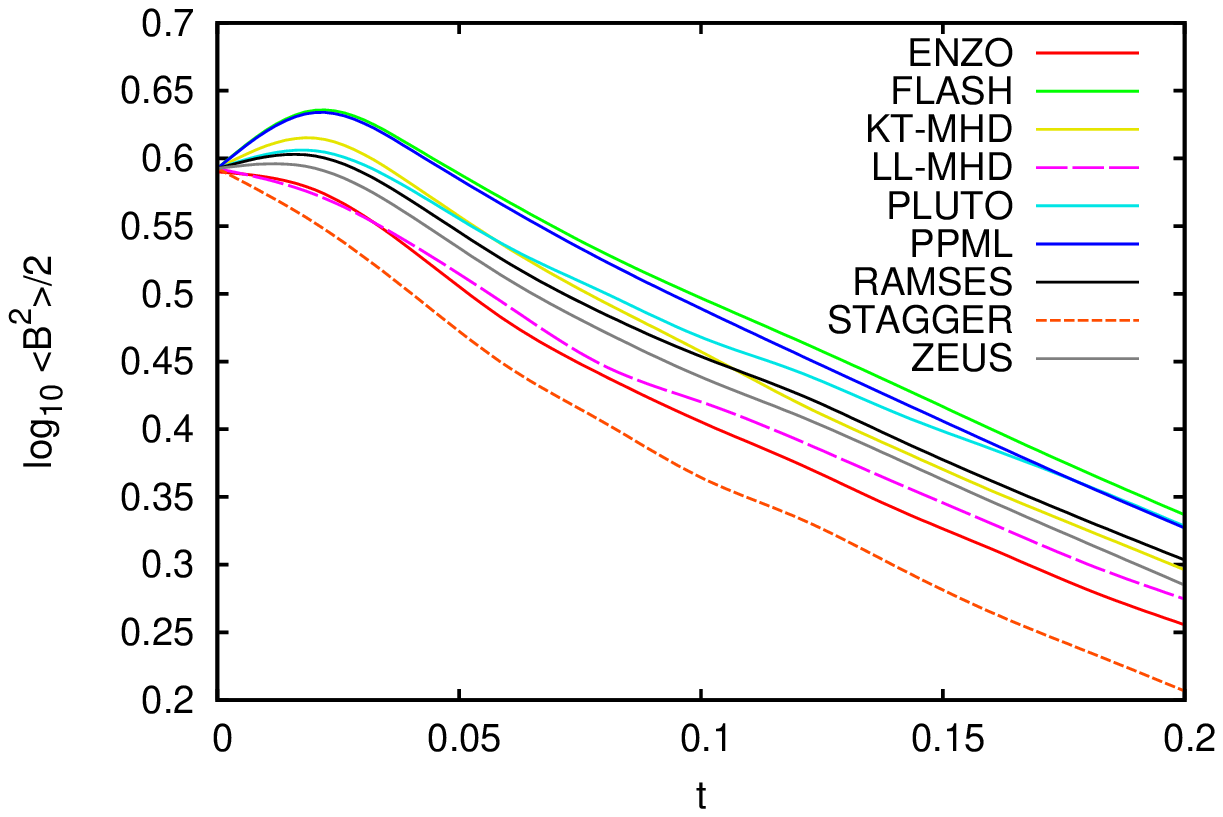}}
\centerline{\plottwo{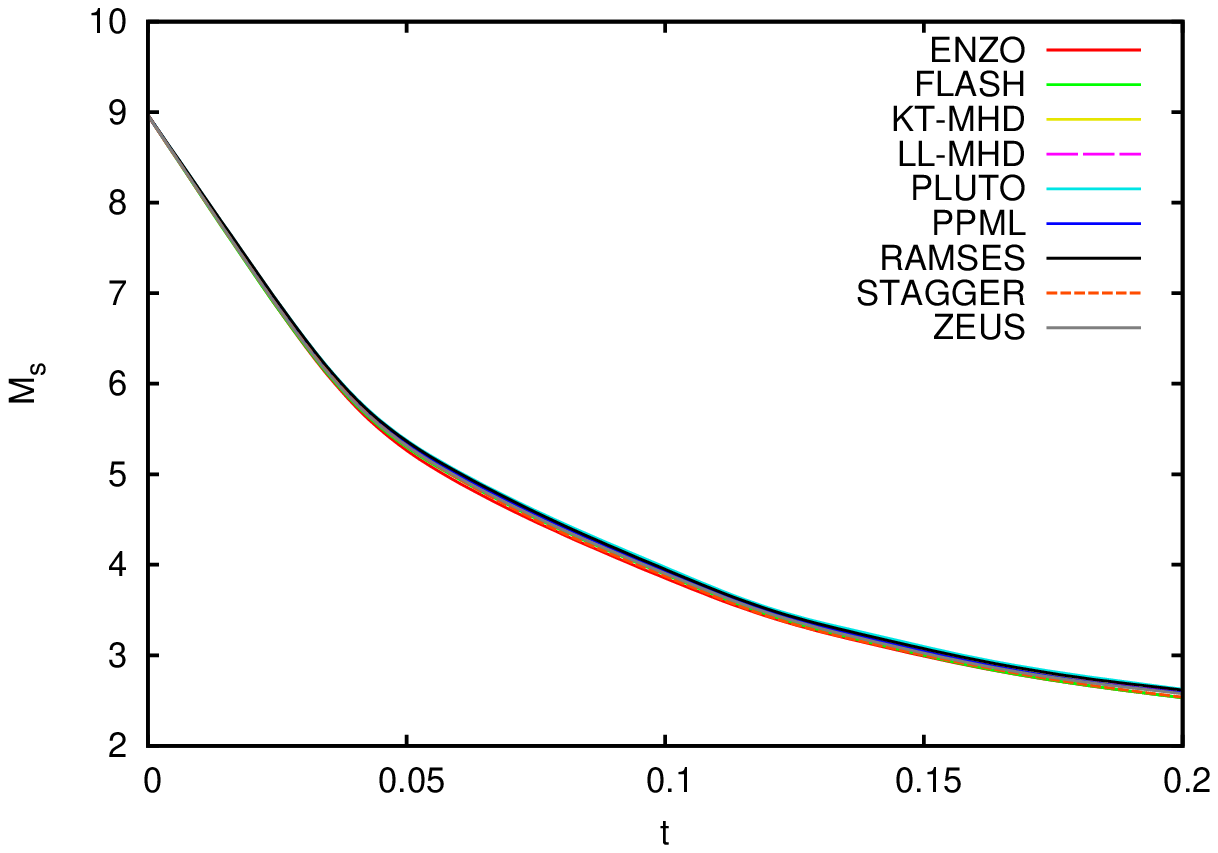}{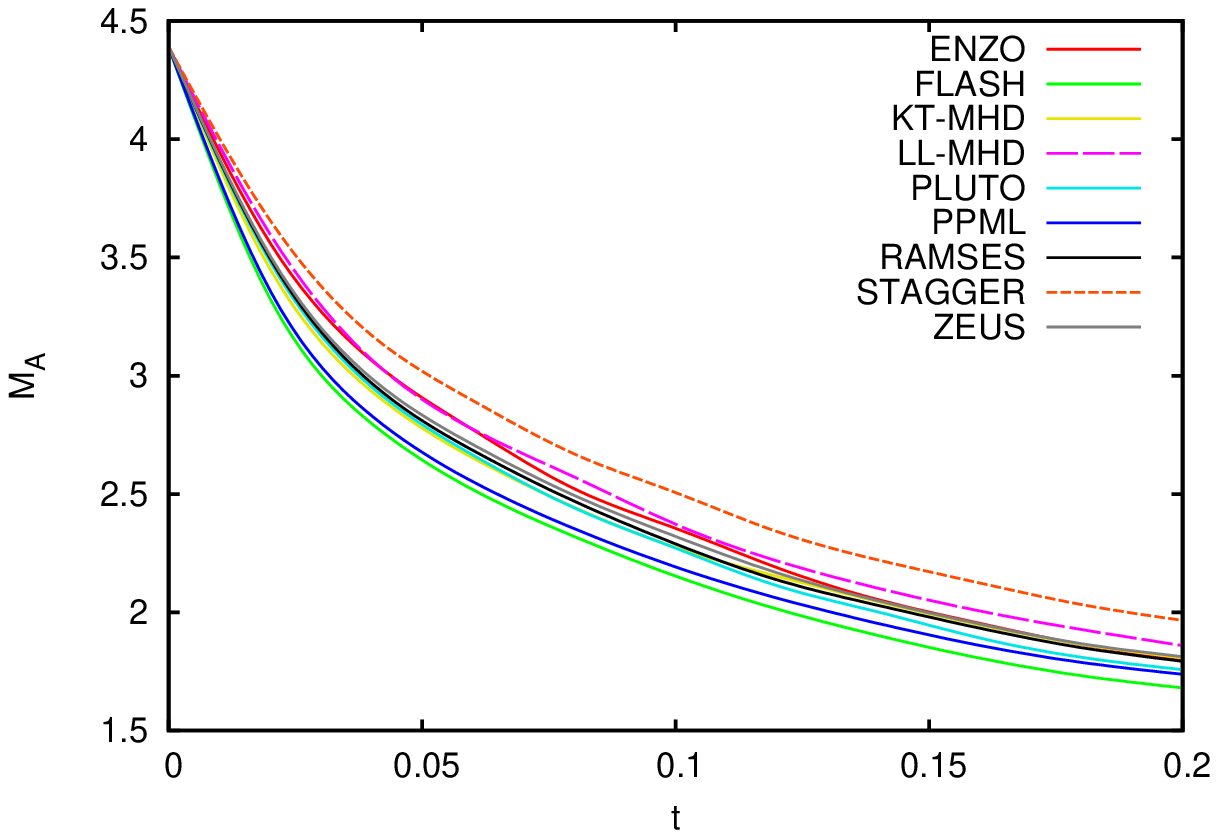}}
\caption{Time evolution of the mean specific kinetic energy (top left), magnetic energy (top right),
as well as sonic (bottom left), and Alfv\'enic (bottom right) rms Mach numbers at grid resolution of $512^3$
cells.
Note, that the kinetic energy and sonic Mach number are rather insensitive to the details of numerical dissipation,
while the evolution of magnetic energy and Alfv\'enic Mach number display significant dependence on the numerical 
magnetic diffusivity.}
\label{energy}
\end{figure*}

\begin{figure*}
\epsscale{1.17}
\plottwo{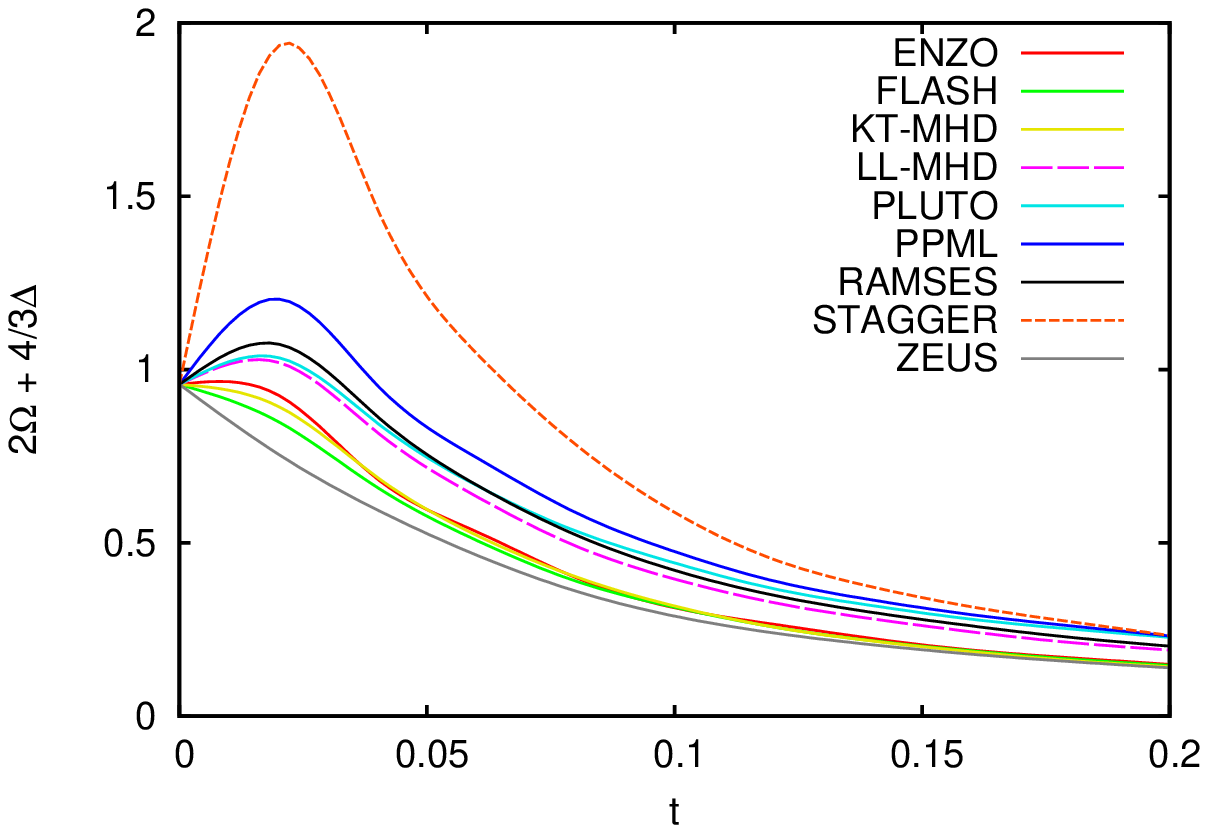}{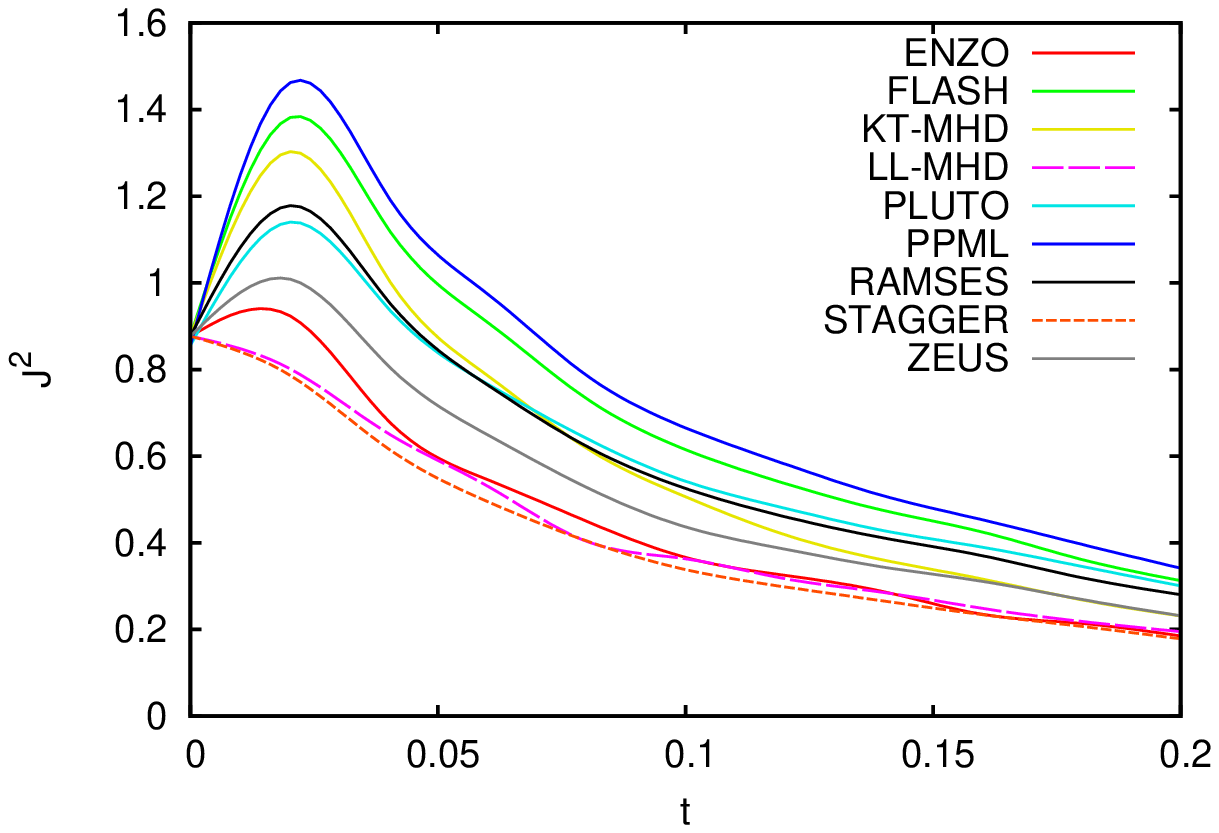}
\caption{Time evolution of $-\epsilon_{\rm K} \mathrm{Re}_{\rm eff}=2\Omega+4/3\Delta$ defined in Section~\ref{scurl} (left) and 
the mean-squared current, $J^2=\left<|\nabla\times {\bf B}|^2\right>$ (right).
These ``small-scale'' measures of turbulent fluctuations are sensitive to the details of numerical
diffusivity and highlight differences between the methods.}
\label{curl}
\end{figure*}

\vspace{.5cm}
\section{Results}
\label{results}
There is a great variety of interesting statistical measures in magnetized 
supersonic turbulent flows to study and compare. The KITP07 project originally 
implied a comparison of density structures in physical space using projections 
and slices, PDFs of the gas density, various power spectra and structure functions, 
time evolution of some global and local average quantities, etc.  Most of these 
data provided by individual contributors to the project can be accessed
electronically via the wiki Web site at KITP under the rubric {\em Star Formation Test Problems}.%
\footnote{http://kitpstarformation07.wikispaces.com/Star+Formation+Test+Problems}

In this paper we mainly focus on the statistics of the basic
MHD fields (the so-called primitive variables) since those are easier to interpret
and link back to the essential features of the numerical methods.
Since the system of equations and the initial and boundary conditions are the
same for all codes, the only source of differences in the numerical solutions is
numerical dissipation. In this section, 
we discuss the sensitivity of various turbulence diagnostics to the numerics and 
describe a set of statistical measures that allow us to assess the quality of 
different algorithms we compare.
We begin with global averages over the periodic domain and then continue with 
analysis of power spectra.
We avoid discussion of density statistics, even though these are important for 
numerical star formation studies. That discussion would be more appropriate in a context
of driven turbulence, where time-averages over many flow snapshots help to
reduce the strong statistical noise associated with the density 
\citep[e.g.,][]{kritsuk..06,kritsuk...07}. The density statistics
in a similar context have been discussed in detail elsewhere \citep[e.g.,][]{kitsionas+12-08,price.10}.

\subsection{Mean Kinetic and Magnetic Energy, rms Mach Numbers}
The evolution of the mean kinetic energy is captured perfectly well
by all methods, except for some small ($<8$\% by $t=0.2$, see Table~1) differences that become 
noticeable at $t>0.12$ in Figure~\ref{energy}. 
This particular quantity is known to converge rather early in nonmagnetized compressible
turbulence simulations, and the same is true for super-Alfv\'enic turbulence 
\citep{lemaster.09,kritsuk...09b}. The velocity power spectrum $P({\bf u}, k)\sim k^{\alpha}$ 
has an inertial range slope $\alpha\in[-2,-5/3]$ that depends on the sonic Mach number 
(see, for instance, Figure~\ref{spectra} below). Because the spectral slope is so steep, the 
mean specific kinetic energy density, $E_{\rm K}\equiv\left<{\bf u}^2/2\right>=\int_0^{\infty} 
P({\bf u},k) d k/2$, is strongly dominated by large scales. 
If the resolution is sufficient to properly capture the structure of large scale flow in the computational 
domain, the energy convergence is achieved. This is apparently the case in our $512^3$ simulations, 
see the left panels of Figure~\ref{energy}. For the $256^3$ model (not shown), the result is very 
similar. We thus conclude that in the decaying turbulence problem the mean kinetic energy is not 
sensitive to variations in small-scale numerical diffusivity between different methods.

The mean magnetic energy density, $E_{\rm M}\equiv\left<{\bf B}^2/2\right>=\int_0^{\infty} P({\bf B},k) d k/2$, 
appears to be more sensitive to variations in small-scale numerical kinetic and magnetic 
diffusivity in super-Alfv\'enic simulations, see Figure~\ref{energy}, right panels, and Table~1. Most of the
methods show an early-time increase in magnetic energy, but asymptotically, after saturation is 
reached, all of them show very similar decay rates $\dot{E}_{\rm M}/E_{\rm M}$. 
The saturated level of $E_{\rm M}$ in the initial flow snapshot generated with the original, non-conservative
version of the STAGGER code is lower than most other codes would produce, except for perhaps LL-MHD and ENZO. 
To compensate for this deficiency of magnetic energy in the initial conditions, FLASH and PPML add about 
$7$\% to the initial $E_{\rm M}$ by the time $t_1=0.02$, when the first flow snapshot is recorded. 
KT-MHD increases $E_{\rm M}$ by  $\sim4$\%, PLUTO, ZEUS and RAMSES add $\sim1$\%$-2$\%. The level of 
$E_{\rm M}$ reached with the old STAGGER code is roughly consistent with that of ENZO and LL-MHD. 
The new, conservative STAGGER appears to be more diffusive than all other methods as far as 
the magnetic energy density is concerned.

To understand why the magnetic energy levels are different, one needs to recall that the convergence 
rate for $E_{\rm M}$ with the grid resolution is rather slow at $512^3$. For instance, with PPML, the 
saturated levels of $E_{\rm M}$ in driven simulations continuously grow as grid resolution improves and 
the convergence is expected only at $2048^3$ or even higher \citep{kritsuk...09b,jones...11}.%
\footnote{\label{iles}
Note that, strictly speaking, the ILES approach involved here does not imply convergence as grid 
resolution improves since the effective Reynolds number is ultimately a function of the grid size 
\citep[e.g.,][]{kritsuk..06}. At the same time, an asymptotic regime corresponding to $\mathrm{Re}=\infty$ can 
potentially be reached relatively early, at large, but still finite Reynolds numbers. This is what 
we probably observe as grid resolution approaches $2048^3$ in super-Alfv\'enic simulations.}
The slow convergence of $E_{\rm M}$ in super-Alfv\'enic runs is not surprising because of their rather 
shallow magnetic energy spectra (see Figure~\ref{spectra} below and note that the magnetic spectra 
are plotted noncompensated).
In such weakly magnetized ($B_0\ll b_{\rm rms}$) isotropic nonhelical flows, turbulence amplifies the 
rms magnetic field fluctuations by stretching and tangling the field lines primarily on small scales 
until an equilibrium is reached between the rms field amplification and dissipation. The saturated level 
of $E_{\rm M}$ for a given sonic Mach number, $M_{\rm s}$, and mean magnetic field strength, $B_0$, would 
naturally depend on the effective magnetic Prandtl number and on the effective magnetic Reynolds number.
We discuss the relative standing of the methods in terms of $\mathrm{Rm}$ and $\mathrm{Pm}$ in the next section and show
that the saturation level of magnetic energy indeed correlates with $\mathrm{Rm}$ and $\mathrm{Pm}$. 
Overall, by $t=0.2$, the deviations of $E_{\rm M}$ from the reference solution defined in Section~5.4
can be as large as 30\%, see Table~1.

Turbulence regimes simulated with different methods differ slightly in their rms Alfv\'enic Mach numbers, 
$M_{\rm A}=\sqrt{2\left<\rho u^2\right>/\left<B^2\right>}$. We use this proxy for the Alfv\'enic Mach number
instead of $\sqrt{2\left<\rho u^2/B^2\right>}$ since in the latter case the locations where $B$ (nearly) 
vanishes would produce arbitrarily large contributions to the mean making this measure unstable. 
%This diagnostic includes a mixture of the gas density, velocity, and magnetic field strength.
%In the middle of the run a

At $t=0.1$, the least super-Alfv\'enic regime is achieved with FLASH and PPML, followed by PLUTO,
KT-MHD, RAMSES, ZEUS, ENZO, LL-MHD, and the new STAGGER in the order of increasing $M_{\rm A}$. 
Note, that the ranking of the methods is essentially the same as for $E_{\rm M}$, which, unlike
$M_{\rm A}$, does not depend on the gas density. This similarity can be explained by a limited 
sensitivity of our chosen proxy for $M_{\rm A}$ to correlations between density and velocity
or between density and field strength \citep{kritsuk...09a}.

Finally note that while $E_{\rm K}$ decays by a factor $>10$ during the course of the simulation, the 
decay of $E_{\rm M}$ proceeds much slower, only by a factor of $\sim2$. Similar to the 
incompressible case \citep{biskamp03}, in supersonic turbulence the energy ratio $\Gamma\equiv E_{\rm K}/E_{\rm M}$ 
is not constant but decreases with time. While $M_{\rm s}$ decreases by a factor of $\sim3$ from $\sim9$ to 
$\sim2.6$, $M_{\rm A}$ shows a $2.5\times$ drop from $\sim4.5$ to $\sim1.8$. These differences in the 
decay rates of kinetic and magnetic energy as well as in the behavior of $M_{\rm s}$ and $M_{\rm A}$ 
can be understood as consequences of self-organization, i.e., the relaxation of the turbulence towards 
an asymptotic static force-free minimum-energy state \citep[e.g.,][]{biskamp03}.

\begin{figure*}
\epsscale{1.16}
\centerline{\plottwo{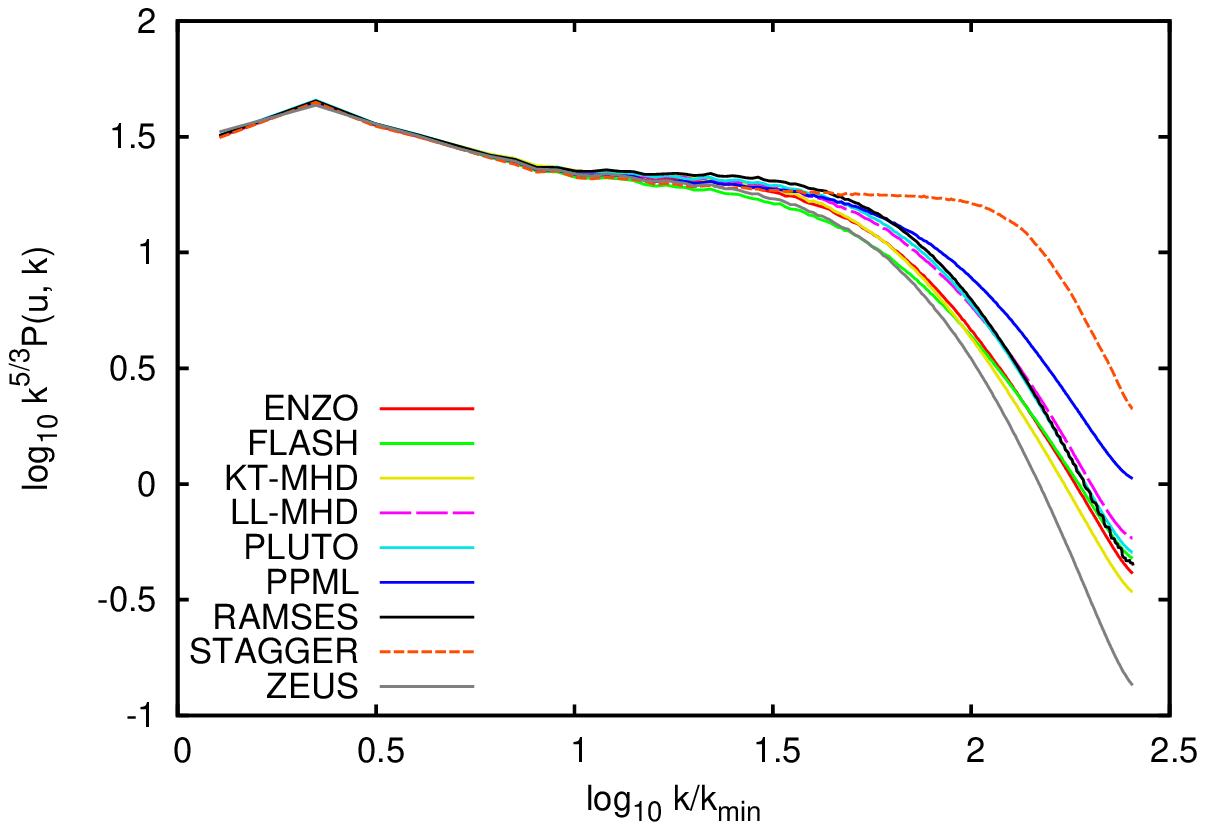}{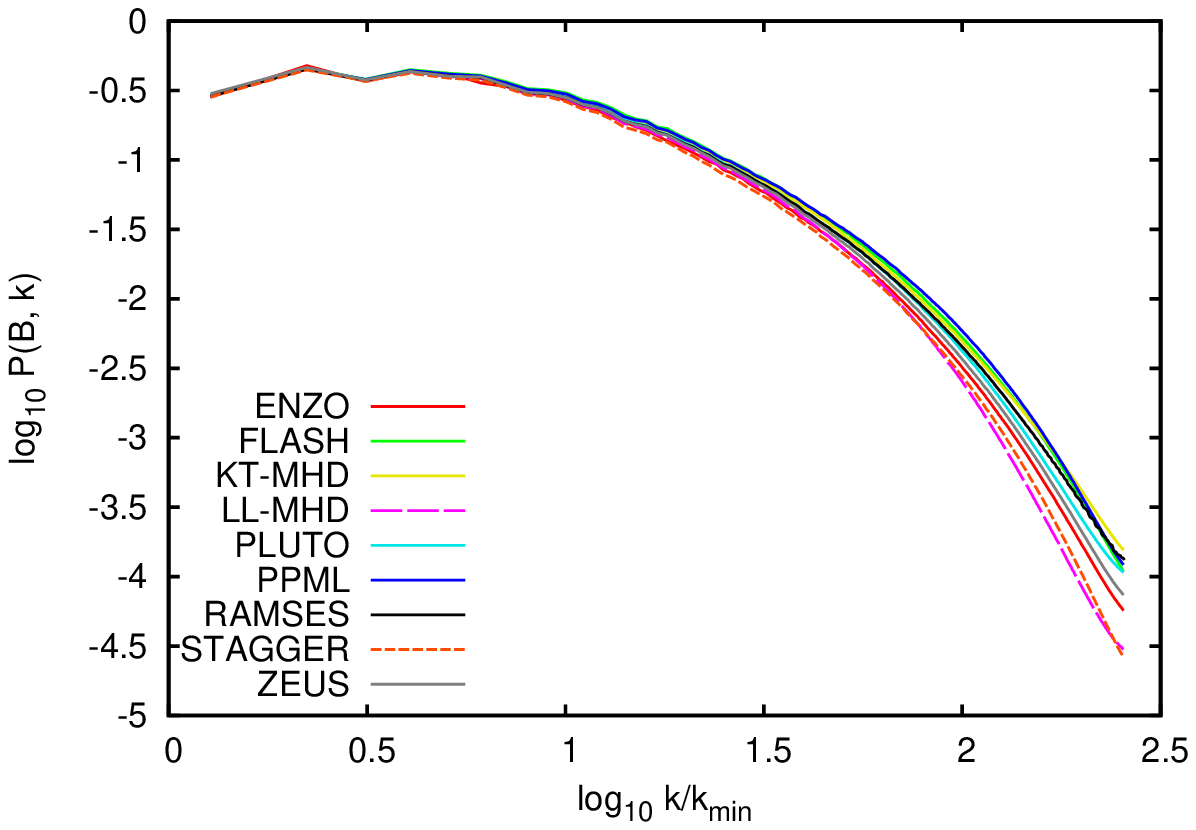}}
\centerline{\plottwo{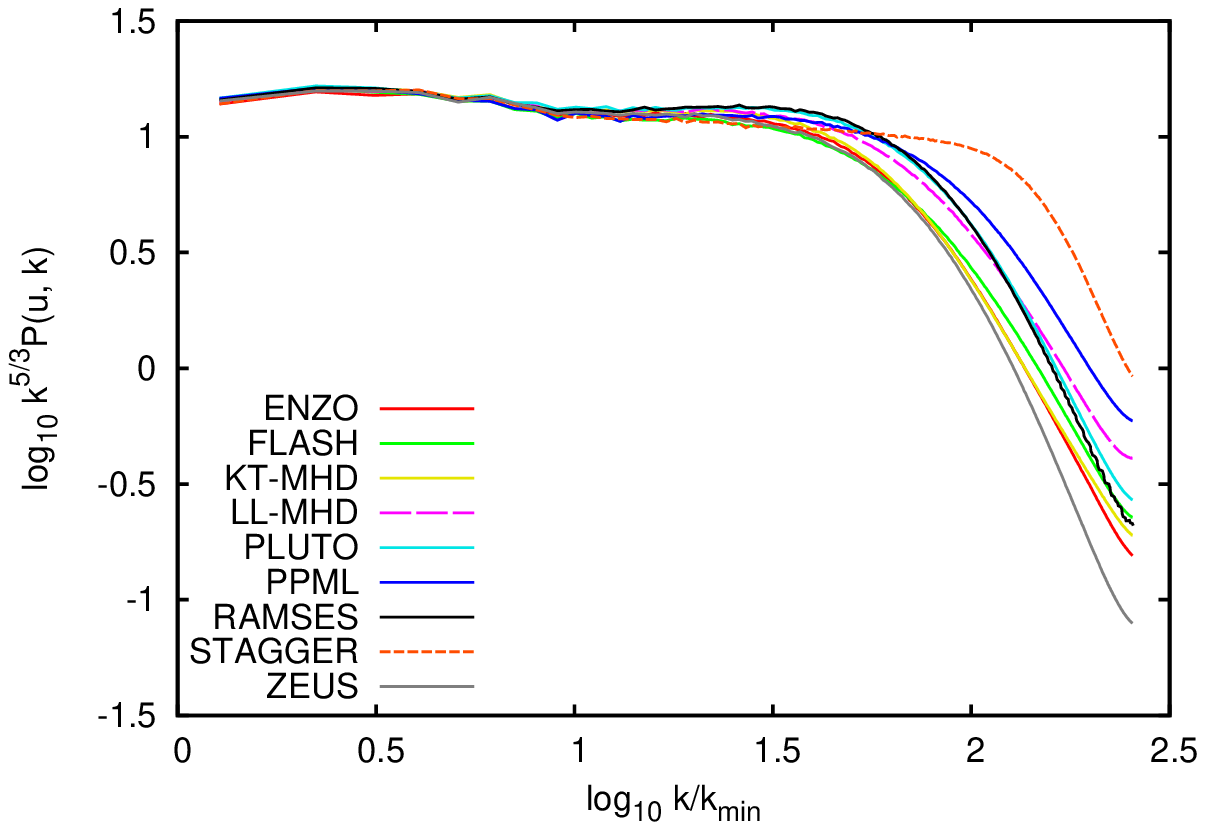}{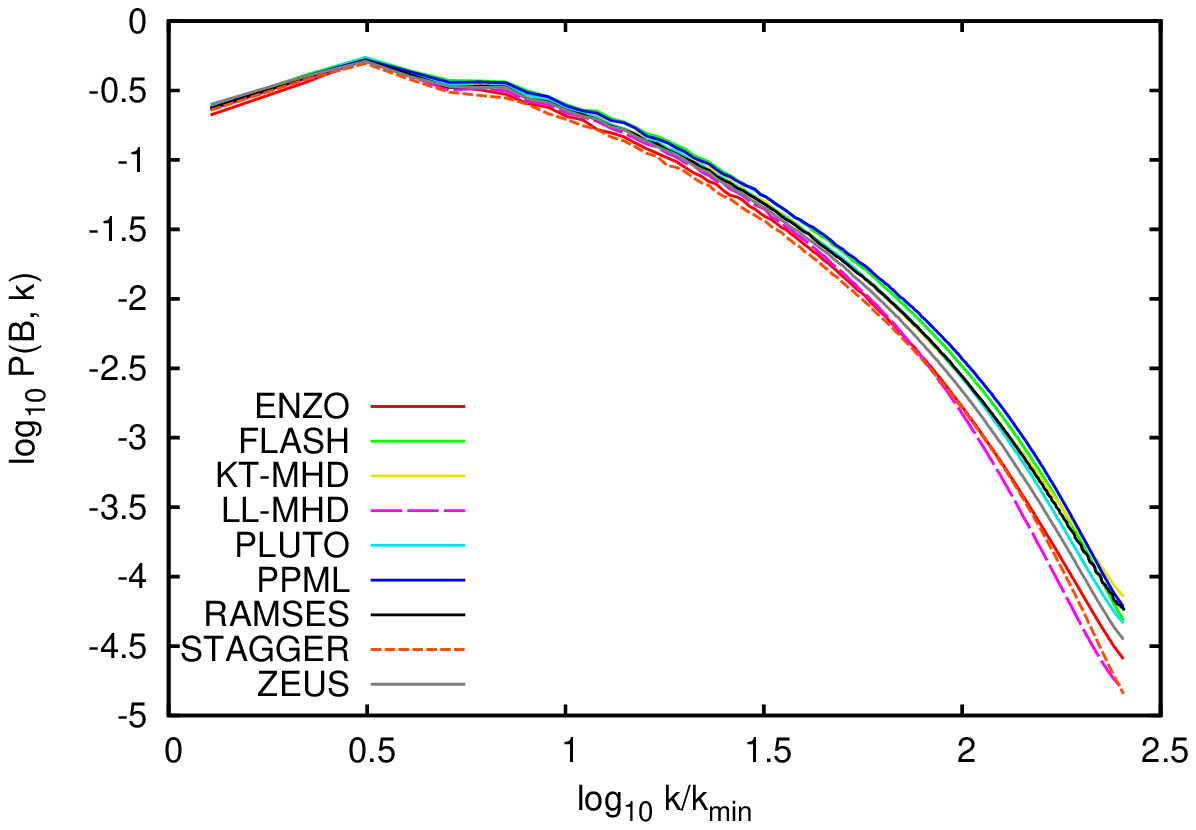}}
\centerline{\plottwo{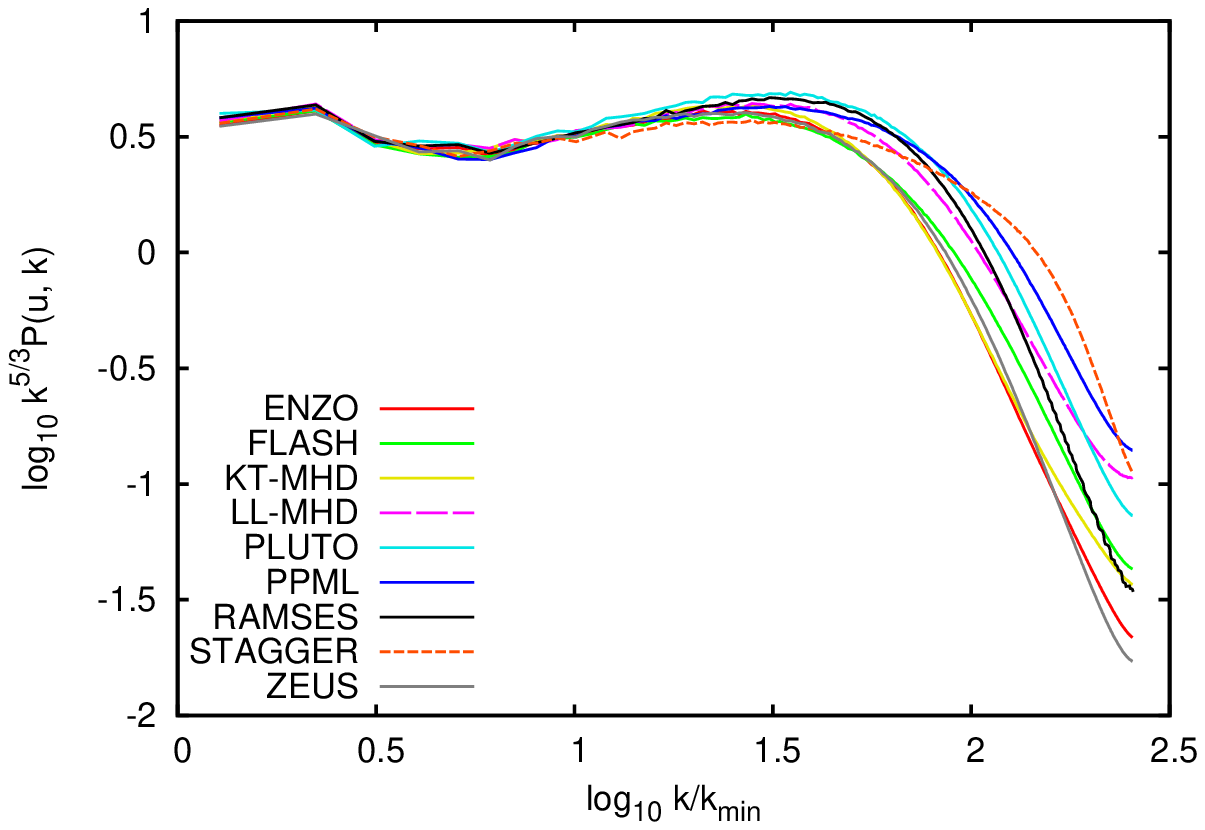}{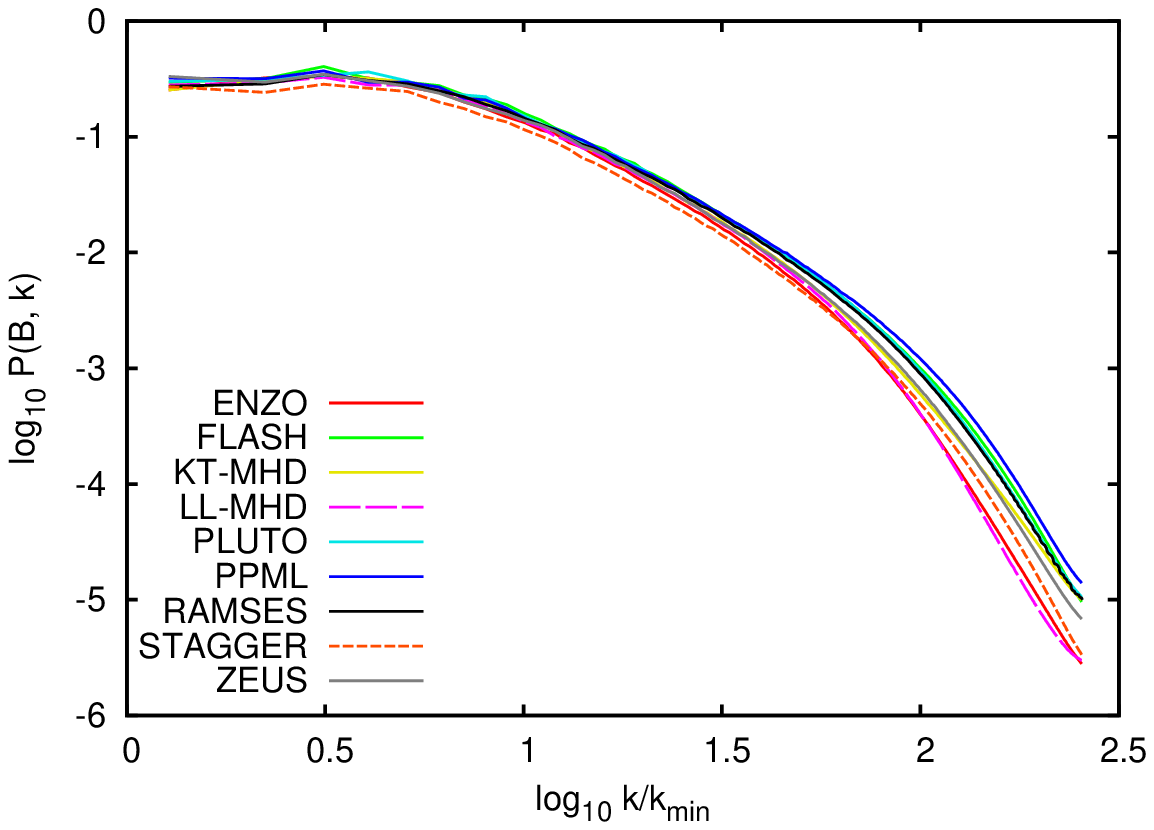}}
\caption{Power spectra of the velocity (left panels) and magnetic energy (right panels) 
on a $512^3$ grid for flow snapshots 1, 3, and 10 at $t=0.02$, 0.06, and 0.2 (top-to-bottom), 
respectively. The velocity spectra are compensated with $k^{5/3}$, while there is no compensation for 
the magnetic energy spectra. Note that the ordinate scale is not always the same for different time 
instances.}
\label{spectra}
\end{figure*}

\begin{figure*}
\epsscale{1.17}
\plottwo{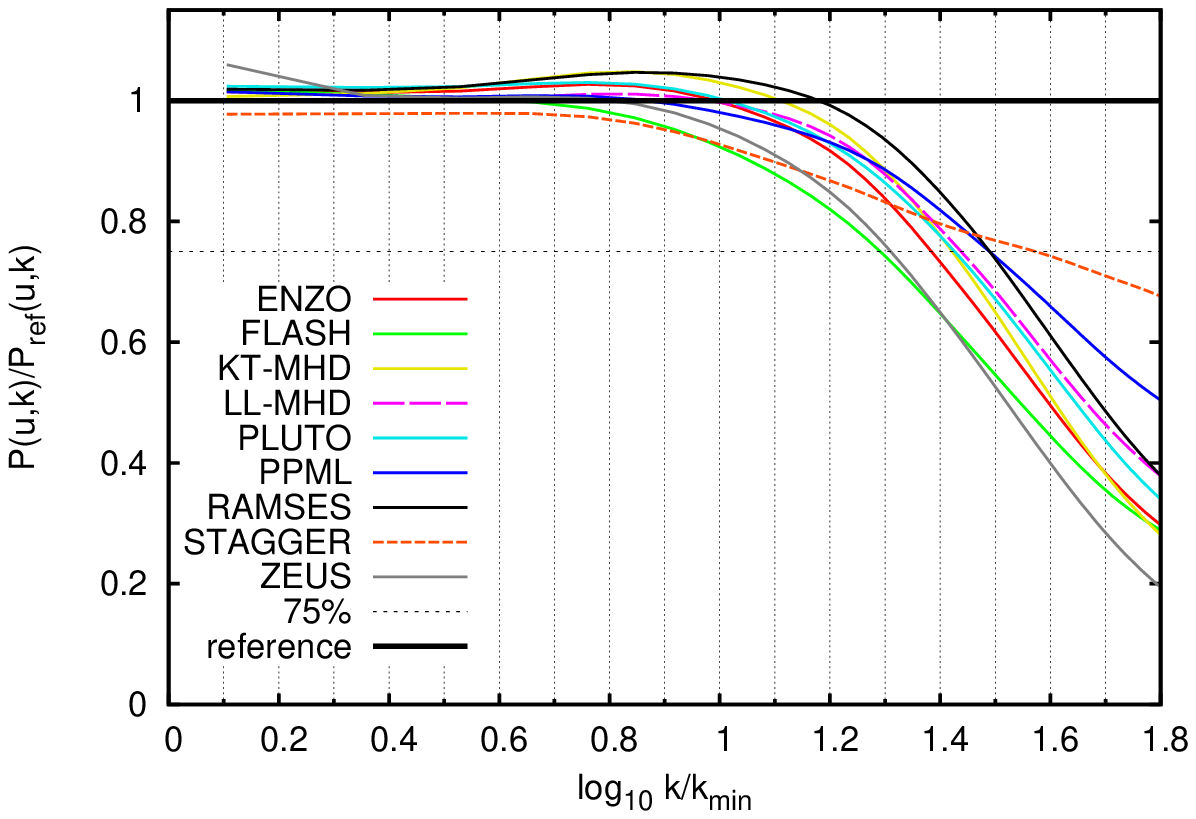}{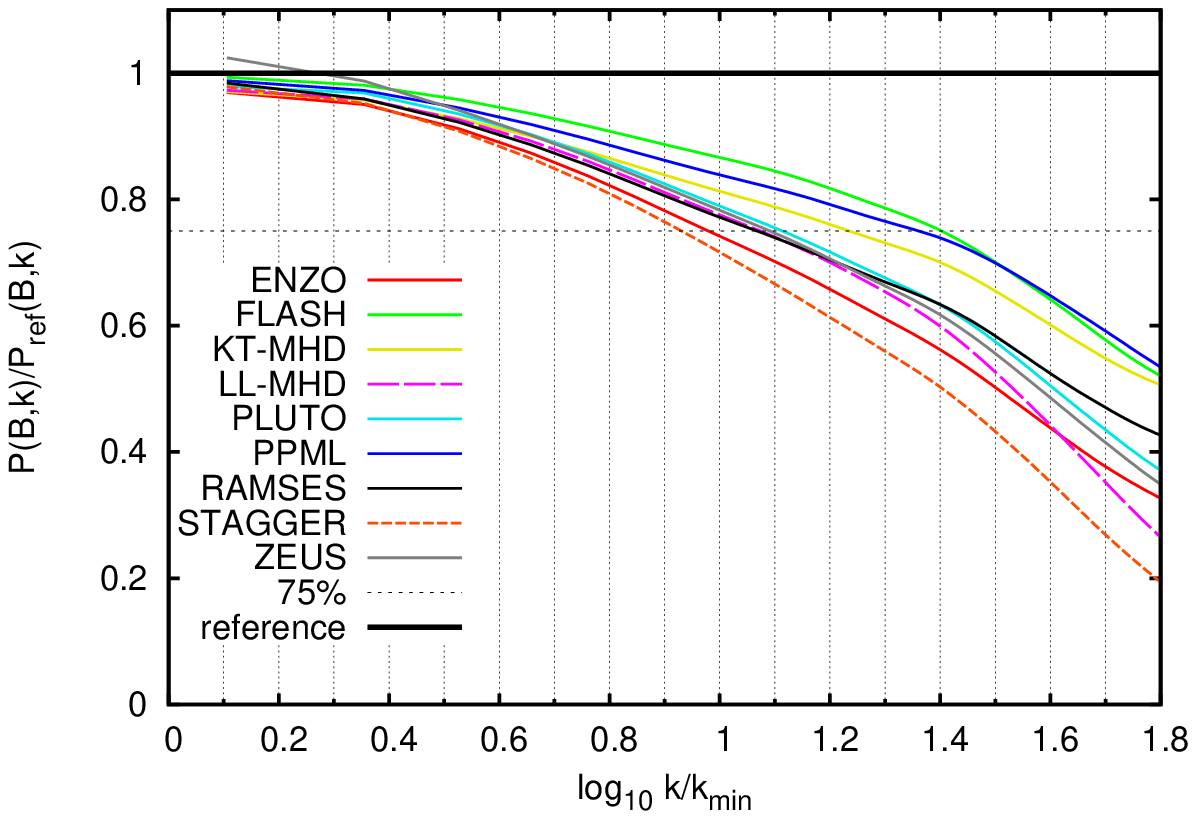}
\caption{Compensated power spectra of the velocity (left panel) and magnetic energy (right 
panel) for the first flow snapshot at $t=0.02$ from the $256^3$ simulations.}
\label{band}
\end{figure*}

\begin{figure*}
\epsscale{1.17}
\plottwo{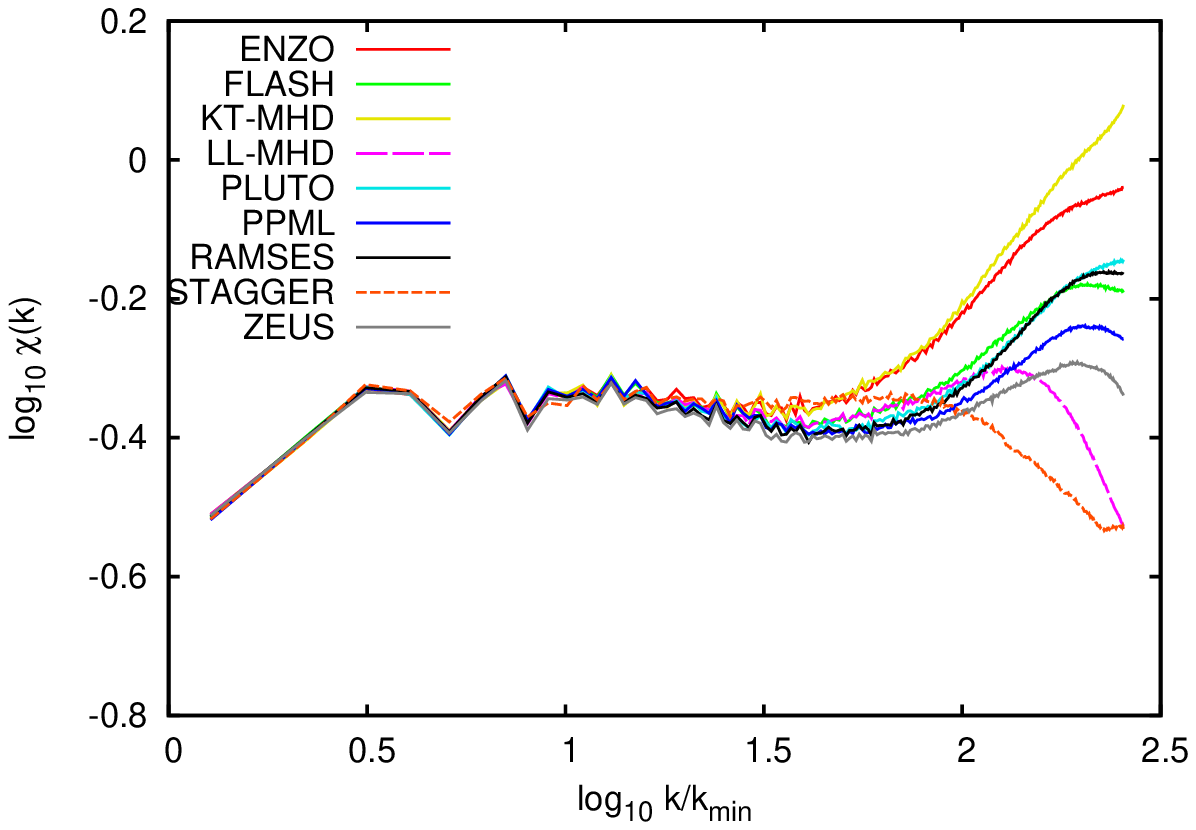}{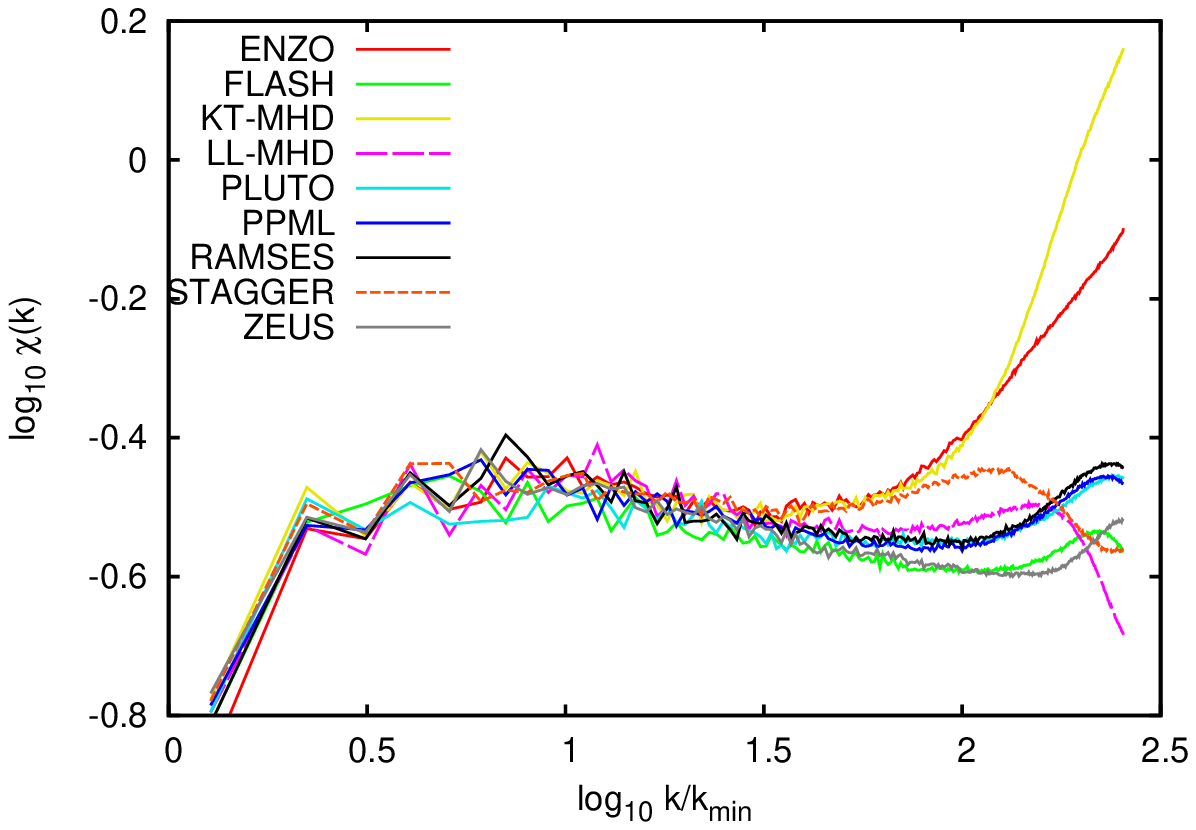}
\caption{Ratio of dilatational-to-solenoidal power in velocity spectra, $\chi(k)$, 
for the first (left panel) and the last (right panel) flow snapshots
from the $512^3$ simulations at $t_1=0.02$ and $t_{10}=0.2$, respectively.}
\label{chi}
%\vspace{1cm}
\end{figure*}

\subsection{Small-scale Kinetic and Magnetic Diagnostics}
\label{scurl}
Another way to look at the effects of numerical dissipation is to analyze the time evolution of enstrophy 
$\Omega=\frac{1}{2}\left<|\nabla\times {\bf u}|^2\right>$, dilatation 
$\Delta=\left<|\nabla\cdot {\bf u}|^2\right>$, and that of the mean squared current density 
$J^2=\left<|\nabla\times {\bf B}|^2\right>$. These so-called ``small-scale'' quantities show strong 
oscillations at the grid scale. Their spectra are dominated by the high wavenumbers, and their PDFs
have extended exponential tails \citep[e.g.,][]{porter..02}. They also usually display a very slow (if any) 
convergence with grid resolution in ILESs due to a strong dependence on $\mathrm{Re}$ and $\mathrm{Rm}$. These measures 
are related to the total viscous and Ohmic dissipation rates within the periodic domain \citep[e.g.,][]{kritsuk...07}. 
For instance, in nonmagnetized compressible turbulence, which is in many respects similar to the 
super-Alfv\'enic case considered here, the mean dissipation rate of the specific kinetic energy 
can be expressed as $\epsilon_{\rm K}=-(\mathrm{Re})^{-1}(2\Omega+4/3\Delta)$ \citep[e.g.,][]{pan..09}.\footnote{%
Strictly speaking, this is only valid for compressible Navier--Stokes turbulence assuming zero bulk viscosity,
i.e., for ideal monoatomic gases.}
Since the global dissipation rates for $E_{\rm K}$ are very similar for all the methods considered here
(see Figure~\ref{energy}), the relative ranking of their effective Reynolds numbers is fully determined 
by the value of $2\Omega+4/3\Delta$. We can thus use the dissipation rates plotted in the left panel of 
Figure~\ref{curl} to determine the relative standing of these methods in terms of their $\mathrm{Re}_{\rm eff}$. 
The new STAGGER code shows an outstanding result during the first half of the evolution, $t\in[0,0.1]$, when
its $\mathrm{Re}_{\rm eff}$ exceeds that of PPML by up to a factor of $\sim1.5$. RAMSES, PLUTO, LL-MHD, ENZO, 
KT-MHD, FLASH, and ZEUS follow STAGGER and PPML in the order of decreasing effective Reynolds number. See Table~1
for numeric values of $2\Omega+4/3\Delta$ from different codes at $t=0.02$.

We employ the same approach to get an assessment of the relative standing of these numerical methods in 
terms of their effective magnetic Reynolds number, $\mathrm{Rm}_{\rm eff}$. Figure~\ref{curl}, right panel 
shows the mean-squared current density, $J^2$, as a function of time. The current density is sensitive 
to both $\epsilon_{\rm M}$ and $\mathrm{Rm}_{\rm eff}$, since $\epsilon_{\rm M}\sim-\left(\mathrm{Rm}\right)^{-1}J^2$. 
We expect qualitatively the same dependency here as for $\epsilon_{\rm K}$ and $\mathrm{Re}_{\rm eff}$, but with, 
perhaps, different order of the methods. Note, however, that the dissipation rates and saturated levels 
of magnetic energy are not the same for different methods as can be seen in Figure~\ref{energy}, right panel. 
PPML shows the highest $\mathrm{Rm}_{\rm eff}$, followed by FLASH, KT-MHD, RAMSES, PLUTO, ZEUS, ENZO, LL-MHD, and STAGGER. Note that 
the order in which methods follow each other in the right panel of Figure~\ref{curl} is the same as in 
the $E_{\rm M}$ plot in Figure~\ref{energy}, see also Table~1. 
Thus, $\mathrm{Rm}_{\rm eff}$ and $E_{\rm M}$ are positively correlated. 
We will explore this correlation further in Sections~\ref{sspectra} and \ref{sband}, where we analyze 
the power spectra of kinetic and magnetic energy and measure the effective bandwidth of the methods.

\subsection{Power Spectra}
\label{sspectra}
Figure~\ref{spectra} shows power spectra of the velocity and magnetic energy at $t=0.02$, 0.06, and 0.2. 
The spectra obtained at $512^3$ demonstrate a very good agreement with each 
other up to $\log_{10}{k/k_{\rm min}}\sim1.2$ and slightly diverge at higher wavenumbers. This means
that numerical dissipation strongly affects the scales smaller or equal to $\sim16$ grid cells in supersonic
turbulence simulations with our best methods.

The new STAGGER
shows a very extended spectrum at $t=0.02$ with an asymptotic slope of $-5/3$ up to $\log_{10}{k/k_{\rm min}}\lsim2$.
This slope is not preserved, however, for the whole duration of the simulation. By $t=0.2$, when after many
integration time steps the sonic 
Mach number drops to $\sim2.5$, the spectrum looses some high-$k$ power and progressively bends down at 
$\log_{10}{k/k_{\rm min}}\gsim1.5$ leaving behind PLUTO, RAMSES, PPML, and LL-MHD.
A close inspection of the velocity spectra shows that numerical diffusion in ZEUS, FLASH, KT-MHD, and ENZO is 
somewhat stronger than in the rest of the grid-based codes and affects the velocities at lower wavenumbers. 
Besides the new STAGGER, RAMSES, PLUTO, PPML, and LL-MHD are the least diffusive codes. The magnetic 
energy spectra display similar variations, but the ranking of methods is different. Here FLASH and PPML 
show very low magnetic diffusivity, while RAMSES, ZEUS, STAGGER, LL-MHD, and ENZO are more diffusive and KT-MHD 
stays in between.

\subsection{Effective Spectral Bandwidth}
\label{sband}
In order to highlight variations in the power spectra obtained with different methods, we follow the
procedure developed in \citet{lele...09} and \citet{johnsen...........10} for compressible Navier-Stokes turbulence
at moderate Mach numbers. We have seen that PPML is one of the least diffusive methods here, 
so we declare the $1024^3$ PPML 
solution filtered with a low-pass Gaussian filter down to $256^3$ to be ``exact'' and call it the 
{\em reference solution}. We then plot power spectra compensated by the spectrum of this reference 
solution for the first snapshot at $t=0.02$, see Figure~\ref{band}. We set tolerance at a level of 
$\pm 25$\% from the reference solution and define the spectral bandwidth of a method as
the fraction of the Nyquist frequency where the compensated spectrum deviates by more than $25$\% 
from the reference solution. While this definition is rather arbitrary, it helps to establish a 
convenient quantitative measure to assert the performance of numerical methods for the turbulence 
decay test, see Table~1 for numeric values.

The left panel of Figure~\ref{band} shows the compensated velocity spectra. STAGGER, PPML, and RAMSES 
have the highest spectral bandwidth in velocity. 
The second position is shared by LL-MHD, PLUTO, KT-MHD, and ENZO. 
ZEUS and FLASH show a similar velocity bandwidth.
RAMSES and KT-MHD show a small bump at $\log_{10}{k/k_{\rm min}}\in[0.6, 1.1]$ reminiscent of the bottleneck effect, while the 
STAGGER and PPML spectra decrease monotonically. 

The right panel of Figure~\ref{band} shows the compensated magnetic energy spectra. The situation here is quite different.
First, unlike the velocity spectra, the magnetic energy spectra start to bend down from the reference solution rather early.
This is expected because of the slow convergence of $E_{\rm M}$ with grid resolution, as we discussed earlier.
FLASH and PPML demonstrate the highest bandwidth, KT-MHD is in the middle, PLUTO, ZEUS, LL-MHD, ENZO, and RAMSES are very similar 
to each other and show a somewhat higher magnetic diffusivity. 
At $256^3$, the spectral bandwidth of our best MHD codes is $\sim0.3$ for the 
velocities and $\sim0.2$ for the magnetic energy. If we were dealing with properly converged solutions obtained from direct
numerical simulations as 
in \citet{johnsen...........10}, that would mean that numerical dissipation strongly affects the wavenumbers down to at 
least $(0.2$--$0.3)k_{\rm N}$. This is not however exactly the case here due to 
the adopted ILES approach, see footnote~\ref{iles}.

\subsection{Dilatational versus Solenoidal Modes}
We have discussed above how the numerical methods differ in their kinetic and magnetic diffusivity. This aspect plays an
important role in simulations involving small-scale turbulent dynamo. There $\mathrm{Pm}$ serves as a control parameter
and the dynamo would only operate at $\mathrm{Pm}>\mathrm{Pm}_{\rm crit}$ \citep{brandenburg.09}. 

In this section, we look at how different methods treat dilatational and solenoidal components of the velocity field on 
small scales. We decompose the velocity fields into the potential (curl-free) and rotational (solenoidal) components, 
${\bf u}={\bf u}_{\rm d}+{\bf u}_{\rm s}$, using Helmholtz decomposition. We compute power spectra, $P({\bf u}_{\rm d},k)$ 
and $P({\bf u}_{\rm s},k)$, and define the dilatational-to-solenoidal ratio as $\chi(k)\equiv P({\bf u}_{\rm d},k)/P({\bf u}_{\rm s},k)$. 
Peculiarities in the small-scale $\chi(k)$ ratio have a potential to affect various turbulence statistics (e.g., the density PDF) 
and limit (or even eliminate) the extent of the inertial range in simulations. These features cannot be captured by either the 
small-scale kinetic diagnostics discussed in Section~\ref{scurl} or by the power spectra discussed in Section~\ref{sspectra}. 
We present $\chi(k)$ for snapshots 1 and 10 ($t=0.02$ and $0.2$) in the left and right panels of Figure~\ref{chi}, respectively.
Table~1 gives numeric values for the average dilatational-to-solenoidal ratio, $\bar{\chi}(k/k_{\rm min}>100)$, at wavenumbers 
above $100k_{\rm min}$ in the $512^3$ models at $t=0.2$.

First, note a very good agreement between all the methods in the early snapshot at low wavenumbers, $\log_{10}k/k_{\rm min}<1.3$, with 
$\chi(k/k_{\rm min}=10)\approx0.47$. An overall ratio of $1:2$ is expected for super-Alfv\'enic 
turbulence at high Mach numbers \citep{kritsuk...09c,federrath....10}. As the turbulence decays, 
the sonic Mach number drops down to $M_{\rm s}\sim2.7$ by $t=0.2$ and the ratio decreases to 
$\chi(k/k_{\rm min}=10)\approx0.33$, as expected. In the inertial range, $\chi(k)$ is known to be
a slowly decreasing function of the wavenumber \citep{kritsuk...07,kritsuk...09c} and this behavior 
is nicely captured by most of the codes. There are slight differences in the $\chi(k)$ levels 
between different methods at $t=0.2$ with ZEUS and FLASH being slightly ahead of the other codes 
in damping the dilatational modes at high wavenumbers. Otherwise, the results from different methods 
look very similar, although ENZO, KT-MHD, and STAGGER start to deviate somewhat from
the rest of the codes at relatively low wavenumbers, $\log_{10}k/k_{\rm min}\sim1.5$.
Also KT-MHD and ENZO produce unusually high $\chi$ values at the Nyquist wavenumber.
For instance, KT-MHD has $\chi(k_{\rm N})\approx1.25$ and $1.5$ at $t=0.02$ and $0.2$, respectively. 
This indicates that perhaps some spurious compressible fluctuations are present at scales 
$k\gsim k_{\rm N}/8$ in simulations carried out with these codes. The small-scale oscillations 
of the KT-MHD code are likely to be caused by the way the CT scheme is 
implemented (R. Kissmann, 2010, private communication). The observed compressible 
artifacts can probably be reduced to a large extent by using  the CT approach proposed by
\citet{londrillo.04}.

\section{Discussion}

\def\divb{\ensuremath{{\bf \nabla\cdot B}=0}}
\def\div{\ensuremath{{\bf \nabla\cdot B}}}

\def\spmhd{NDSPMHD}

One is tempted to try to sort these nine methods into some well-ordered set.
This is, however, an impossible task, as no single solver consistently
outperformed all others on all diagnostics.  In this discussion we will restrict our focus to
discussing kinematic and magnetic dissipation,  as measured by the diagnostics
presented here.  This leaves out other potentially important diagnostics, such
as the loop advection test of \citet{2005JCoPh.205..509G}.  We are also ignoring other salient features, such as computational
cost (in memory and time to solution), ease or feasibility of extending the
solver to different physical or numerical scenarios, or quality of
documentation, all of which go into the selection of a code package.  The final
result is that all codes performed reasonably well on the task. There is no
single silver bullet that determines the performance of a given solver; good
quality results can be achieved through a variety of means, and dissipation can
be introduced in a variety of places.

All MHD algorithms used in this work are extensions of a
previously established hydrodynamic algorithm.  In general, five basic features determine the operation
of a numerical scheme; base method (most prominently spatial order of accuracy), MHD
extension, artificial 
viscosity, time integration, and directional splitting.  In this section, we will
classify each code based on these parameters and discuss trends within each
feature.  In Section~\ref{sec.spatial}, we discuss the spatial order of
accuracy, which seems to be the dominant factor in determining performance.  In
Section \ref{sec.viscosity}, we will discuss artificial viscosity and source
terms that behave as a viscosity. These features also seem to have
considerable impact on the dissipation properties, as one would expect.  In
Section \ref{sec.base}, we will classify and discuss other properties of the base
hydrodynamical scheme.  In Section \ref{sec.mhd}, we will discuss the performance of
various MHD extension methods.  In Section \ref{sec.three}, we will discuss three closely
related schemes and discuss how seemingly small details can dramatically effect
the results.  In Section \ref{sec.others} we will discuss directional splitting,
time integration, and reconstructed variables.  These seem to have a less
dramatic impact on the overall performance, as measured by these diagnostics.
See Table~2 for a summary of these solver configuration details.

We refer the reader
to the excellent books by \citet{Toro}, \citet{Laney98}, and 
\citet{Leveque02} and the method papers cited
in Section 3 for the details of each numerical scheme.  We will not be expanding
on any details except where necessary.

\ctable[star,pos=t, caption = Solver Design Specifications for the Eulerian Methods$^{\rm a}$]{l l c c l l l}{
\tnote[{\rm a}]{See Section \ref{sec.methods} and the indicated sections on each topic for more information.}
\tnote[{\rm b}]{Base method. FD for finite difference, FV for finite volume.  FV
techniques have the Riemann solver listed.  Section
\ref{sec.base}.}
\tnote[{\rm c}]{Spatial order of accuracy, Section \ref{sec.spatial}. }
\tnote[{\rm d}]{Artificial Viscosity, Section \ref{sec.viscosity}. ``$\shortparallel$ Derivative'' indicates
presence of terms proportional to the longitudinal derivative of the magnetic field.}
\tnote[{\rm e}]{MHD method, Section \ref{sec.mhd}.}
\tnote[{\rm f}]{Time integration method, Section \ref{sec.time}.}
\tnote[{\rm g}]{Multidimensional technique, Section \ref{sec.split}. ``$\perp$ Reconstruction'' 
indicates presence of transverse derivatives in the interface reconstruction.}
}{
\hline
\hline
Name & Base Scheme\tmark[{\rm b}]& Spatial Order\tmark[{\rm c}] & Source Terms\tmark[{\rm d}]& MHD\tmark[{\rm e}] & Time Integration\tmark[{\rm f}] &
Directional Splitting\tmark[{\rm g}]   \\
\hline
\ENZO   & FV, HLL      & 2nd & Dedner                      & Dedner        & 2nd-order RK    & Direct                 \\
\FLASH  & FV, HLLD     & 2nd & $\shortparallel$ Derivative & 3rd-order CT  & Forward Euler   & $\perp$ Reconstruction \\
\KTMHD  & FD, CWENO    & 3rd & KT                          & 3rd-order CT  & 4th-order RK    & Direct                 \\
\LLMHD  & FV, HLLD     & 2nd & None                        & Athena CT     & Forward Euler   & Split                  \\
\PLUTO  & FV, HLLD     & 3rd & Powell                      & Powell        & 3rd-order RK    & Direct                 \\
\PPML   & FV, HLLD     & 3rd & None                        & Athena CT     & Forward Euler   & $\perp$ Reconstruction \\
\RAMSES & FV, HLLD     & 2nd & None                        & 2D HLLD CT    & Forward Euler   & $\perp$ Reconstruction \\
\STAGGER& FD, Stagger  & 6th & Tensor                      & Staggered CT  & 3rd-order Hyman & Direct                 \\
\ZEUS   & FD, van Leer & 2nd & von Neumann                 & MOC-CT        & Forward Euler   & Split                  \\
\hline
}

\subsection{Spatial Order of Accuracy}\label{sec.spatial}

High spatial order of accuracy seems to be the salient feature of the least
dissipative codes, though as there are many factors in each method that can
improve or degrade performance.  STAGGER has the highest spatial order, 6, and
this is reflected most notably in Figure~2, left panel, where its effective Reynolds number
is significantly higher than the other methods, and Figure~3, left panel, where the
inertial range of the power spectrum extends much further than the others.
The third-order methods are PPML,
PLUTO, KT-MHD, and the electric field construction of FLASH.  These four methods
show the highest magnetic spectral bandwidth, and are the top performers in the
effective magnetic Reynolds number and magnetic power spectrum.  However, other
effects, most likely viscosity, keep these third-order methods from having 
the lowest dissipation among all statistics.  The remaining methods (ZEUS,
RAMSES, LL-MHD, ENZO, and the base hydro scheme of FLASH) are
second order spatially.  These codes tend to show more dissipation over the third-order
methods.  

There are two notable exceptions to this trend.  The first can be seen
in the spectral bandwidth plot, in
which RAMSES (second order spatially) performs better than some third-order
methods, though this may be due to other effects (see Section \ref{sec.three}).
The second exception can be seen in the top curve of the effective Reynolds
number, corresponding to the STAGGER method.  The initial conditions were
generated with an early version of STAGGER, but continued with a version that
used conservative variables and different settings for the artificial viscosity.
As both methods are sixth order spatially, the increase in effective Reynolds
number demonstrates that it is not spatial accuracy alone that determines
dissipation properties.  This will be discussed further in Section~\ref{sec.viscosity}.
Note that here we refer only to the formal spatial order of accuracy
for the reconstruction or interpolation for each scheme.  The actual
convergence properties of each scheme, once time integration, spatial
reconstruction, etc. have been taken into account, must be measured as
a function of time and/or space resolution.  This is beyond the scope
of this work.

\vspace{.5cm}
\subsection{Artificial Viscosity and Source Terms}\label{sec.viscosity}
It is quite typical for numerical schemes to include some form of artificial
viscosity in order to avoid numerical instabilities. In the case of the \citet{Powell99} and
\citet{dedner.....02} MHD schemes, source terms proportional to \div\ are included to
constrain the effects of divergence, which while not the same kind of
dissipation still have a dissipative effect.  In this suite of simulations,
viscosity treatments can be broken coarsely into three categories:  artificial
viscosity, \div\ motivated diffusivity, and exclusively numerical viscosity.
Explicit terms are included in STAGGER, KT-MHD, ZEUS. Terms due to \div\
treatments are included in PLUTO and ENZO.  The
four remaining codes have no explicit artificial viscosity, and dissipation is 
only due to the scheme itself (these are FLASH, PPML, RAMSES, and LL-MHD).

One naively expects that codes with explicit viscosity terms will have somewhat
more dissipation than those without.  However, this is only loosely seen in 
the results, and it is difficult to disentangle dissipative terms from other code
differences.  STAGGER gives the most noticeable example of the effects of
dissipation, namely the large gap between its velocity dissipation, which is
quite low, and its magnetic dissipation, which is quite a bit higher than other
codes on most metrics.  It is also possible that the fine tuning of the
magnetic and kinematic artificial diffusivity, which has maximized the apparent inertial range, has
altered the non-local coupling of MHD waves in a manner that still leaves the
dissipation relatively high in the inertial range.  It is reasonable to isolate codes based on spatial order
of accuracy in order to compare viscosity results.  Among the third-order codes, PPML, with no explicit
viscosity, tends to have lower dissipation than PLUTO, which has \divb\
motivated source terms, which in turn tends to have lower dissipation than
KT-MHD.  Then isolating the second order methods, the trend somewhat continues,
though less robustly.  ENZO and ZEUS tend to show the most velocity dissipation,
as measured by effective Reynolds number or velocity bandwidth.  However, ENZO
is the only second spatial order code with explicit magnetic dissipation, and it
shows more power in the magnetic power spectrum than LL-MHD, which has none.
RAMSES  shows the lowest dissipation among the second order codes in
all metrics except for magnetic spectral bandwidth, in which ZEUS is slightly
higher.

One method (FLASH) includes terms proportional to the longitudinal
derivative of the magnetic field.  These terms are typically omitted
from the derivation as they are identically zero in the one-dimensional 
version of the equations.

\subsection{Base Methods} \label{sec.base}

Eulerian hydro schemes fall broadly into two categories: finite-volume and
finite difference.  In loosest terms, finite volume schemes approximate the
integral form of the conservation law, while finite difference terms approximate
the differential form of the conservation law.  Three of the grid-based codes compared here are
finite difference: STAGGER, KT-MHD, and ZEUS.  The other six are finite-volume methods
(ENZO, FLASH, LL-MHD, PLUTO, PPML, and RAMSES.)
Between finite volume and finite-difference, there is no correlation with
performance.  This is best illustrated in the left panel of Figure~2.  The code
with the highest effective Reynolds number is STAGGER, and with the lowest is
ZEUS, and both are finite difference methods.   Here we will discuss some
common features within each category of methods.

\subsubsection{Finite-difference Methods}

One of the curses of numerical fluid dynamics is the battle between accuracy and
stability.  These seem to be felt somewhat more strongly by the three finite-difference 
codes.  ZEUS tends to be more dissipative than other methods, even
though it is formally second order spatially (see, for instance, the
effective Reynolds number in Figure~2, left, or the left column of Figure~3).
STAGGER has the highest effective fluid Reynolds number, but the lowest effective
magnetic Reynolds number; we believe this to be a result of the tensor viscosity
and its subtle relationship to the (scalar) magnetic diffusivity. The KT-MHD method
suffers from excessive small-scale compression, likely due to the fact that CWENO schemes
are only {\it essentially} non-oscillatory, trading the possibility of small
numerical oscillation near shocks for very high quality results in smooth
regions. 

\subsubsection{Finite-volume Methods}

The six finite-volume methods 
(ENZO, FLASH, LL-MHD, PLUTO, PPML, and RAMSES)
are all some form of higher-order  extension of
Godunov's method.  These methods have the advantage that they capture shock
structures, in principle, exactly.  These methods can be broken into two parts;
reconstruction to the interface, and the Riemann solver.   

A wide array of
Riemann solvers exist in the literature, but those used in this work are of
two families, Roe and HLL.  It is expected from other tests that Roe will
perform the best, though it is subject to instabilities at low density and high
Mach numbers, and HLLD will perform the best of the HLL methods, as it captures
more of the eigenstructure of the equations than HLL.
However, there does not seem to be a correlation between
dissipation and choice of Riemann solver that cannot be explained by other
mechanism.  This does not say that there is no difference, merely not one that
can be identified by these data.  

The interface reconstruction techniques vary widely among the six schemes, and
can primarily be characterized by details discussed in other sections.  Namely,
order of reconstruction, directional splitting, time integration, and explicit
viscosity terms.  They will not be discussed further here.

\vspace{.5cm}
\subsection{MHD Methods}\label{sec.mhd}
Any MHD algorithm is essentially an established hydrodynamic algorithm with modifications
to include the Lorentz force in the momentum equation, the induction equation,
and some treatment to minimize the divergence of the magnetic field.
In all cases in this paper, the Lorentz force is incorporated into the momentum
equation directly (rather than through, say, vector potentials).  
%A few (ENZO and PLUTO) incorporate
%explicit sources into the momentum update to deal with the reality that \div\
%is non-zero numerically that do serve as source terms to the momentum equation
%as well as the induction and energy equations.  
Two of the codes (ENZO and PLUTO) use
non-exact divergence preservation, namely both treat an extra wave, in \div.
These methods also include source terms to Equations (1)--(3) that are
set to zero in most methods.
This extra wave is advected and damped in ENZO, while it
is simply advected with the fluid velocity in PLUTO and serves to
mitigate singularities in the three-dimensional linearized Jacobian.
The rest use a variant of the CT method, wherein the electric field and magnetic field are treated at
the zone edge and face, respectively, which allows the solenoidal constraint to
be kept zero to machine precision through the curl operator.

One naively expects the two approximate divergence methods to have somewhat
higher dissipation than other codes, as the primary driver is dissipation.  This
is to some extent seen in the data, though
PLUTO does not suffer much from this effect as it still has quite high 
fluid and magnetic Reynolds numbers.  ENZO, on the other hand, seems to
have more dissipation, and based on its similarity to other spatially second-order 
codes, the \div\ wave seems to be a likely culprit.

Among the CT-based schemes, the results seem to be dominated by first
spatial order then on reconstructed variable. 
PPML and LL-MHD both use the electric field reconstruction technique
described in the Athena method of \citet{2005JCoPh.205..509G}, but PPML is
spatially third-order, so has a higher magnetic Reynolds number.  It
should be noted that only the electric field reconstruction of the
Athena method is used by these methods. FLASH also uses third-order reconstruction,
and also has an extremely large magnetic spectral bandwidth.  LL-MHD,
ZEUS, and RAMSES, on the other hand, are all spatially second-order,
but ZEUS uses MOC, which uses the characteristic fields, and RAMSES
solves a second Riemann problem; both of which prove to better capture
the electric field than the primitive variable reconstruction used in
LL-MHD.

\subsection{Three Closely Related Codes}\label{sec.three}

\def\kmax{\ensuremath{\log_{10} k/k_{\rm min}}}

An interesting subset of codes to examine are LL-MHD, RAMSES, and FLASH.  These
three codes are the most similar in terms of their components, and serve to
illustrate how small differences in method details can cause significant
differences in the performance.  Each of the three codes uses the
second-order MUSCL-Hancock reconstruction-evolution scheme for computation of
interface states, the HLLD Riemann solver, forward Euler time
integration, and a higher-order CT method.  Given all the similarities, the
differences in performance of the three codes are surprising.  
This is best shown in the two spectral bandwidth plots in Figure~4.  The magnetic
bandwidth of FLASH, in Figure~4, right, is the highest of all available codes, as
measured by the wavenumber at which the spectrum crosses 75\%, with
\kmax=1.4.  RAMSES and LL-MHD are significantly lower, both with
\kmax=1.1.  The spatial order of accuracy in the electric field computation
is a clear culprit.  FLASH uses a
third-order central-difference reconstruction of the electric fields from the
Riemann solver.  RAMSES and LL-MHD, on the other hand, both use spatially second-order 
methods, with RAMSES using a novel two-dimensional Riemann solver, and
LL-MHD using the Athena method.  This shows the importance of spatial
reconstruction in capturing flow features.  The velocity bandwidth, in Figure~4, 
left, is a completely different story:  RAMSES is at the high end of the
codes, with \kmax=1.5, LL-MHD is in the middle, with \kmax=1.4, but
FLASH is the lowest of the Eulerian codes, with \kmax=1.1.  This is most
interesting, since the base solvers for each of the codes are nearly identical.
The biggest difference here is the treatment of directional splitting.  RAMSES
and FLASH are both directionally unsplit, incorporating transverse derivatives
of the Jacobian in the interface reconstruction as discussed in Section
\ref{sec.split}, while LL-MHD is split using Strang
splitting, and does not include transverse derivatives.  This alone does not
explain the ordering, as LL-MHD and RAMSES perform quite similarly in many
velocity statistics.  
The only other major algorithmic difference is the inclusion of the
longitudinal magnetic derivatives in the FLASH interface
reconstruction. It is not obvious that these terms would cause
diffusion in the manner observed, though they will affect the
reconstruction of the interface states.  Finally, each of these
methods (indeed all methods described here) include a number of
nonlinear switches that determine behavior near shocks, among other
things, that have not been explicitly described.  Further
investigation is required to isolate these finer details.

Due to the tight coupling between the velocity and the magnetic field
in both the momentum and induction equations, it would not be
surprising for the velocity and magnetic statistics to be coupled,
even perhaps in an inverse manner, through either energy conservation
or mode coupling. Thus the higher spatial order used in the FLASH
magnetic reconstruction may, for instance, be more efficient at
transferring kinetic energy to magnetic energy.  More study is
required to definitively pinpoint the cause of differences between
these three codes, but it illustrates the effect of seemingly minor
details having substantial results in the behavior of a code.

%The only other difference that might be responsible is the inclusion of
%the longitudinal magnetic derivatives in the FLASH interface reconstruction.  It
%would seem that this term has dissipative effects, as it directly influences the
%reconstruction of the velocity fields.  While not an explicit source term in the
%momentum equation, as discussed in Section \ref{sec.viscosity}, it seems to be
%the only remaining difference to explain the differences between the RAMSES and
%FLASH performance.   However, due to the tight coupling between the velocity and the
%magnetic field in both the momentum and induction equations, it would not be
%surprising for the velocity and magnetic statistics to be coupled,
%even perhaps in an inverse manner, through either energy conservation or mode
%coupling.  More study is required to definitively
%pinpoint the cause of differences between these three codes, but it illustrates the effect of
%seemingly minor details having substantial results in the behavior of a code.

An additional point of interest in the RAMSES behavior is the excess power seen
in the spectral bandwidth plot at $\log_{10} k/k_{\rm min}\in[0.8, 1.2]$.  This
seems to be a manifestation of what in pure hydrodynamic turbulence is referred
to as the bottleneck.  This is typically not seen in simulations of MHD turbulence 
at a $512^3$ resolution, presumably
due to additional effects of non-local MHD mode coupling that allows energy to be
more efficiently transferred to smaller scales.  As RAMSES has a relatively low
Prandtl number, it is possible that this extra energy transfer is not as
efficient as in other codes, causing somewhat inflated spectral bandwidth.

\subsection{Other Solver Details}\label{sec.others}

There are several other solver design specifications that have received
considerable attention over the years.  Here we present a discussion of some of
the major solver options that have been examined.  While each may be crucial in
its own right, are not dominant factors determining the dissipation properties
studied here.

\subsubsection{Evolved Variables}

MHD can be described by three distinct sets of variables: primitive variables
$(\rho, {\bf u}, {\bf B}, p_{\rm{tot}})$; conserved variables $(\rho, \rho
{\bf u}, {\bf B}, E_{\rm{tot}})$, and the
characteristic variables, $R^k$, which are the eigenvectors of the Jacobian of the
equations, and in some ways the most physically relevant form of the variables.
It has been shown that in some cases working with 
the characteristic variables gives superior results to the other two \citep{Balsara04b}.
There is some evidence that bears this out in these data.  For instance, the magnetic
behavior of ZEUS, in which MOC traces characteristics to compute the electric
field, is generally less dissipative than LL-MHD, which uses primarily primitive
variables.  However, this is not universally the case, and other factors may
prove more important.  Such is the case in 
velocity performance of FLASH vsersus RAMSES, which use spatial limiting on
characteristic and primitive variables, respectively.  

\subsubsection{Directional Splitting}\label{sec.split}

Computational algorithms have a long history of being developed as one-dimensional
methods.  They then must be extended by some manner to three dimensions.
There are essentially three categories of multidimensional techniques employed
by the codes in this study; directly unsplit, directionally split, and
transverse flux methods.  

The two directionally split methods (LL-MHD and ZEUS) employ
sequential one-dimensional solves along each coordinate axis, wherein the
partial update of one sweep is used as the initial data for the following sweep.
The order of sweeps in both methods is permuted to reduce error.  In both
methods, the electric field is computed after the three sweeps are finished.

The four ``directly unsplit'' methods are STAGGER,
KT-MHD, PLUTO, and ENZO.  The first two  do not rely on strictly one-dimensional 
techniques, so they employ fully three-dimensional evolution by
repeated application of the interpolation and derivative operators.  The ENZO
and PLUTO methods use the Godunov method, which is strictly speaking one-dimensional as
will be discussed below.  It applies the algorithm in an unsplit fashion,
with the initial state for Riemann solves coming from the same data for all
three dimensions.  It incorporates multidimensional properties of the flow
by way of the Runge--Kutta integration.

Another unsplit technique, dubbed ``transverse reconstruction'' here, is used to
incorporate three-dimensional terms into the finite-volume methods.
Godunov's method is, strictly speaking, one-dimensional, and does not 
lend itself directly to multidimensional techniques. The underlying one-dimensional 
method follows three basic steps;
reconstruction of two interface states at each zone boundary, followed by solution of the
Riemann problem at the zone boundary, and finally using that solution to compute
and difference fluxes at the interface to update the field.  Three of the schemes
(FLASH, PPML, and RAMSES) introduce the multidimensional terms in the
reconstruction of the interface state, through the addition of terms
approximating gradients of the transverse fluxes, $\partial F_y/ \partial y$.  
The techniques vary slightly between the four methods. FLASH and RAMSES use a 
linearization of the transverse flux gradient, $A_y \partial U/ \partial y$ to 
compute the half step advance in time. Here, $A_y$ is the Jacobian of the flux,
and the derivative is approximated with monotonized central differences.  PLUTO
uses a full reconstruction and Riemann solve in the transverse direction.  PPML
also includes a linearization of the transverse flux, though it is incorporated
slightly differently, with the transverse flux gradient introduced in the 
characteristic invariants, and uses characteristic direction filtering for 
upwinding the derivative.

While it is often believed that directional sweeping is a detriment to
the solution quality, the metrics presented in this work do not show a clear
correlation between the different multidimensional techniques. 

\subsubsection{Time Integration}\label{sec.time}
In principle, the order of the spatial and temporal integration should
be the same, otherwise the convergence properties of the scheme will be reduced
to the lower of the two.  However, time integration in these cases seems to be swamped by other effects.  
ENZO is of higher
order in time than RAMSES, but significantly more dissipative.  Similarly, PLUTO
is higher order in time than PPML, but typically has higher dissipation, as
well.  

\section{Summary and Conclusions}
We have compared nine numerical MHD codes on 
a decaying supersonic, super-Alfv\'enic turbulence test problem with 
conditions similar to star-forming molecular clouds in the Galaxy. 
The codes ENZO, FLASH, KT-MHD, LL-MHD, PLUTO, PPML, RAMSES, STAGGER, and
ZEUS, described in detail in Section~2, employ a variety 
of numerical algorithms of varying order of accuracy, multidimensional and 
time integration schemes, shock capturing techniques, and treatment of the 
solenoidal constraint on the magnetic field. Together, they represent a 
majority of the MHD codes in use in numerical astrophysics today and 
therefore sample the current state of the art. The work presented in this 
paper is the largest, most comprehensive MHD code comparison on an 
application-like test problem to date.

The codes were compared using both integrated and spectral measures of 
the velocity and magnetic fields. All nine Eulerian codes agreed surprisingly 
well on the kinetic energy decay rate (Figure~1, top left), which indicates 
both the robustness of published predictions 
\citep{maclow...98,stone..98,lemaster.09} as well as the 
inadequacy of this particular metric as a discriminant among methods. 
All nine Eulerian codes likewise agreed on the magnetic energy decay rate 
(Figure~1, top right), but varied on the amplitude of the peak magnetic 
energy as this proved sensitive to the effective magnetic Reynolds number 
of the simulation, which depends on the numerical dissipation of the method.

To move beyond simple global energy diagnostics, small scale kinetic and 
magnetic field diagnostics were introduced in order to empirically measure 
the effective fluid and magnetic Reynolds numbers of the various codes. 
These diagnostics are based on analytically motivated combinations of the 
volume integrated fluid enstrophy, dilatation, and square of the electric 
current density (Figure~2). They proved more revealing about the numerical 
dissipation present in the various methods, and motivated a closer 
investigation using power spectra of the velocity and magnetic fields. 
Regarding the latter, the concept of effective spectral bandwidth (ESB) was 
introduced as a quantitative metric for code comparison. The effective 
spectral bandwidth is defined as the width in wavenumber space where the 
numerical results do not deviate from a reference solution (typically, 
a higher resolution simulation) by more than 25\%. The ESB was measured for 
both the velocity and magnetic power spectra for all nine codes at reference 
times during the decay. A detailed comparison of ESB's leads to several 
general conclusions and observations. 

\begin{enumerate}
\item All codes gave qualitatively the same results, implying that 
they are all performing reasonably well and are useful for scientific investigations. 

\item No single code outperformed all the others against all metrics, although 
higher-order-accurate methods do better than lower-order-accurate methods in 
general. The lack of a clear winner stems from the fact that a single MHD code 
is a combination of many different algorithms representing specific design choices, 
and that many combinations are possible. 

\item The spatial order of accuracy is the primary determinant of velocity 
spectral bandwidth and effective Reynolds number. Higher spatial order correlates 
to higher spectral bandwidth. The sixth-order code STAGGER is superior to the third-order 
codes PPML, PLUTO, KT-MHD and FLASH, which are superior to the second-order 
codes ZEUS, LL-MHD, and ENZO. 

\item Codes with high velocity spectral bandwidth do not necessarily have high 
magnetic spectral bandwidth. For example, the STAGGER code has the highest 
velocity ESB but the lowest magnetic ESB. The magnetic ESB is sensitive to the 
spatial order of accuracy of the electric field computation, and is higher in 
methods that interpolate on characteristic variables as opposed to primitive variables. 

\item The use of explicit artificial viscosity to stabilize shock waves reduces 
the velocity spectral bandwidth relative to methods that do not use artificial 
viscosity, such as Godunov methods. 

\item The use of explicit divergence cleaning reduces the magnetic spectral 
bandwidth relative to codes that preserve the solenoidal condition on ${\bf B}$ exactly 
(CT methods).

\item Other algorithmic choices such as finite-difference versus finite-volume 
discretization, directionally split versus unsplit updates of the conservations 
laws, and order of accuracy of the time integration are less well correlated 
with the performance metrics, and therefore appear to be less important in 
predicting a code's behavior on MHD turbulence. 

\end{enumerate}

Observations about specific codes are as follows:

\begin{itemize}

\item The best performers overall are PPML, FLASH, PLUTO, and RAMSES based 
on velocity and magnetic Reynolds numbers and spectral bandwidths.
\item The highest fluid Reynolds number was obtained with the STAGGER code.
\item The highest magnetic Prandtl number was obtained with the FLASH code. 
\item FLASH is somewhat more diffusive on the hydro part than its magnetic 
part, and the reverse is true for the RAMSES code.
\item The dilatation velocity power spectra of KT-MHD and ENZO 
exhibit problematic behavior on small scales that is likely related to
the ways these codes maintain \divb.

\end{itemize}

The best performing codes employ a consistently high order of accuracy for 
spatial reconstruction of the evolved fields, transverse gradient 
interpolation, conservation law update step, and Lorentz force computation. 
Three of the four employ divergence-free evolution of the magnetic field 
using the CT method, and all use little to no explicit 
artificial viscosity. These would seem to be guidelines for the development 
of future schemes. Codes which fall short in one or more of these areas are 
still useful, but they must compensate higher numerical dissipation with higher 
numerical resolution. A new class of nearly Lagrangian methods for hydrodynamics has 
recently emerged which uses a moving mesh based on Voronoi cells \citep{springel10}. 
It remains to be seen if this approach can be generalized to MHD while retaining 
the beneficial elements of successful Eulerian schemes.

\acknowledgments
This work was prepared in part during the workshop ``Star Formation Through Cosmic Time'' 
at the KITP in Santa Barbara, and was supported in part by the National Science Foundation 
under grant no. PHY05-51164. Computer support for this project was partly provided by 
the San Diego Supercomputer Center, through an LRAC supercomputer allocation in 
support of the Computational Astrophysics Data Analysis Center.
A.K. was supported in part by the National Science Foundation under
grants AST0507768, AST0607675, AST0808184, and AST0908740.
D.C. was supported in part by the National Science Foundation under
grants AST0808184, and AST0908740. 
D.C. and H.X. were supported in part
by Los Alamos National Laboratory, LLC for the National Nuclear
Security Administration of the US Department of Energy under contract
DE-AC52-06NA25396.
Simulations with ENZO and PPML utilized NSF TeraGrid resources provided by SDSC,
NICS, and TACC through allocation MCA07S014.
\AA .N. was supported in part by the Danish Natural Research Council. STAGGER code results 
were computed at the University of Copenhagen node of the Danish Center for Scientific Computing.
P.P. is supported by MICINN (Spanish Ministry for Science and Innovation) grant AYA2010-16833 
and by the FP7-PEOPLE-2010-RG grant PIRG07-GA-2010-261359.  
R.B. was funded during this research by the Deutsche
Forschungsgemeinschaft under the grant KL1358/4-1.
C.F.~has received funding from the European Research Council under the 
European Community's Seventh Framework Programme (FP7/2007-2013 grant 
agreement no.~247060) for the research presented in this work. The 
FLASH simulations were run at the Leibniz Rechenzentrum (grant pr32lo) 
and the J\"ulich Supercomputing Centre (grant hhd20).
The FLASH code has been developed by the DOE-supported ASC/Alliance
Center for Astrophysical Thermonuclear Flashes at the University of
Chicago. 
D.L. was supported by the U.S. Department of Energy
under grant no. B523820 to the Center for Astrophysical Thermonuclear
Flashes at the University of Chicago.
M.F. thanks Andrea Mignone and Natalia Dzyurkevich for their comments on this
test. The PLUTO calculation was done on the ``Theo'' cluster of MPIA Heidelberg.
P.S.L. is supported by the NASA ATFP grants NNG06-GH96G and NNX09AK31G.
Work of C.V. and W.C.M. was supported by the DFG cluster of excellence: Origin and Structure of 
the Universe. 
The RAMSES simulations were performed thanks to the HPC resources of 
CCRT under the allocations 2009-SAP2191 and 2010-GEN2192 made by GENCI, France.
The authors are grateful to the anonymous referee for a prompt and constructive review.

\end{document}